\documentclass[reprint, aps, onecolumn, notitlepage, pre
]{revtex4-1}
\usepackage{graphicx}
\usepackage{natbib}
\usepackage{amsmath}

\usepackage{bm}
\usepackage{bbm}
\usepackage{xcolor}
\usepackage{transparent}
\usepackage{mathtools} 
\usepackage{float}
\usepackage[colorlinks=true, linkcolor=blue, urlcolor=blue, citecolor=blue]{hyperref}
\expandafter\ifx\csname package@font\endcsname\relax\else
 \expandafter\expandafter
 \expandafter\usepackage
 \expandafter\expandafter
 \expandafter{\csname package@font\endcsname}
\fi
\usepackage{silence} 
\begin{document}
\title[]{
Shear-banding instability in arbitrarily inelastic granular shear flows
}

\author{Priyanka Shukla}
\email[(corresponding author) ]{priyanka@iitm.ac.in}
\homepage{https://home.iitm.ac.in/priyanka/}

\author{Lima Biswas}
\affiliation{Department of Mathematics, 
Indian Institute of Technology Madras, Chennai 600036, India
}

\author{Vinay Kumar Gupta
}
\email{vinay.gupta@warwick.ac.uk}
\homepage{https://warwick.ac.uk/fac/sci/maths/people/staff/gupta/}
\affiliation{
Mathematics Institute, University of Warwick, Coventry CV4 7AL, UK
}
\affiliation{
Department of Mathematics, 
SRM Institute of Science and Technology, Chennai 603203, India}


\begin{abstract}
%
One prototypical instability in granular flows is the shear-banding instability, in which a uniform granular shear flow breaks into alternating bands of dense and dilute clusters of particles having low and high shear (shear stress or shear rate), respectively. 
In this work, the shear-banding instability in an arbitrarily inelastic granular shear flow is analyzed through the linear stability analysis of granular hydrodynamic equations closed with Navier--Stokes-level constitutive relations. 
It is shown that the choice of appropriate 
constitutive relations plays an important role in predicting the shear-banding instability. 
A parametric study is carried out to study the effect of the restitution coefficient, channel width and mean density. 
Two global criteria relating the control parameters are found for the onset of the shear-banding instability.     

\end{abstract}

\maketitle

\section{Introduction}
Granular materials in the so-called rapid flow regime~\cite{Campbell1990,GoldI2003} exhibit various non-uniform structures, such as granular vortices, density waves, clustering, shear banding, etc.~\cite{GNB1996,OK2000,AT2006,ACG2009}.
Among all the non-uniform structures exhibited by granular flows, the shear banding in granular flows---in particular---has received tremendous attention mainly due to its analogy with the shear banding in soft matters, e.g.~foams, emulsions, colloidal suspensions, etc.~\cite{SvH2010}.
The shear banding is manifested even in a granular shear flow, one of the simplest types of flows, which serves as a prototype for the rheology and pattern formation~\cite{LSJC1984,JR1985PoF,SG1998,AL2003b,KM2006,SA2009,
SA2011a,SA2011}.
%
Several experimental (see e.g.~\cite{Mandl1977, VHB1999,Mueth2000,Viggiani2001,BLSLG2001,CG2004, LHLT2010,Murdoch2013}) as well as theoretical studies 
(see e.g.~\cite{AN1998,CG2004,CLG2006,KM2006,CLG2006,
Khain2007,SA2009,
Khain2009, SA2011,SA2011a,BJ2015}) have confirmed the shear-banding phenomenon in granular shear flows.
The theoretical studies based on linear and nonlinear stability analyses~\cite{AN1998,SA2009,SA2011,SA2011a} show that a granular shear flow admits stationary and traveling instabilities leading to clustering and shear banding---with the latter being the focus of the present work.
In the phenomenon of the shear banding, a homogeneous flow transforms into an inhomogeneous flow characterized by the coexisting bands of different rheological properties due to the shearing motion. 
Shearing along the streamwise direction renders inhomogeneities in the other two directions---commonly referred to as the gradient (or transverse) direction and the vorticity (or spanwise) direction. 
Consequently, shearing along the streamwise direction leads to two different banding instabilities, namely the gradient banding instability and the vorticity banding instability.   
In the former, bands of high and low \emph{shear rate} form along the gradient direction while bands of high and low \emph{shear stress} appear along the vorticity direction in the latter. 
The localized high and low shear (shear rate or shear stress) regions correspond to low and high density regions, respectively.  
Both types of the shear-banding instabilities have also been observed in experiments~\cite{Mandl1977,VHB1999, CG2004,Mueth2000,CLG2006} conducted by shearing a granular material in a shear cell wherein shearing remains localized within the narrow regions leaving behind bands of unsheared regions.
For more details, the reader is referred to the articles~\cite{Fielding2007,DB2008}, which provide comprehensive details of the shear-banding phenomenon.

To investigate such a vast variety of phenomena exhibited by granular flows through hydrodynamic theory is one of the most challenging tasks among the granular community. 
The hydrodynamic modeling of granular fluids is more involved in comparison to regular fluids since interactions among granular particles are inherently inelastic; this very nature of granular materials poses some undesirable complexities, e.g.~microscopic irreversibility, lack of scale separation, mesoscopic flow behaviour, strong nonlinearities in the momentum and energy balance equations~\cite{Campbell1990,GoldI2003}. 
Notwithstanding, hydrodynamic models for dilute/dense granular flows can be derived from the (inelastic) Boltzmann/Enskog--Boltzmann equation  within the framework of kinetic theory. 
Similarly to the well-established Navier--Stokes and Fourier (NSF) equations for regular fluids, a hydrodynamic model for granular fluids consists of the mass, momentum and energy balance equations, and the pressure tensor and heat flux appear as additional unknowns in the momentum and energy balance equations. 
However, unlike the NSF equations for regular fluids, the energy balance equation in the case of granular fluids contains an additional term, referred to as the \emph{collisional dissipation}, that accounts for the energy loss due to inelastic collisions among granular particles, and the constitutive relations for the pressure tensor and heat flux are, in general, quite different from the Navier--Stokes' and Fourier's laws for regular fluids.
For granular fluids, the constitutive relations for the pressure tensor, heat flux and collisional dissipation are typically derived from the Boltzmann equation (in the dilute case) or from the Enskog--Boltzmann equation (in the dense case) by means of the Chapman--Enskog expansion at first-order of expansion, see e.g.~\cite{
BDKS1998, GD1999, SG1998, 
BP2004, Gupta2011}. 
Alternatively, these constitutive relations can also be derived from the moment equations, see e.g.~\cite{JR1985, JR1985PoF, KM2011, Garzo2013, GS2017, GST2018}. It is worthwhile to note that determining the constitutive relations from the moment equations is not only much simpler than that by the Chapman--Enskog expansion performed on the full Boltzmann equation but can also yield more accurate constitutive relations on considering more moments \cite{GST2018}.
The mass, momentum and energy balance equations for granular flows closed with the first-order constitutive relations are referred to as the \emph{granular NSF equations}.
The validity of the granular NSF equations---even for rapid granular flows~\cite{Campbell1990,GoldI2003}---is subjected to the conditions under which the constitutive relations for the pressure tensor, heat flux and collisional dissipation are derived. Hence the constitutive relations involved in the granular NSF equations ought to be chosen carefully, especially while dealing with dense granular flows for which the ``molecular chaos'' assumption is inadequate.

The papers~\cite{JR1985, JR1985PoF} by Jenkins and Richman (JR)
may be regarded as the pioneering works which derive the balance equations and associated constitutive relations for dense granular gases of identical rough circular disks and spheres, respectively, based on the revised Enskog theory.
%
%
It is important to note that the restitution coefficient $e$ in the JR model
enters only the energy balance equation through the collisional dissipation while the pressure tensor and heat flux do not depend on $e$ due to the approximations made in their theories. 
Consequently, the NSF transport coefficients---that appear in the constitutive relations for the pressure tensor and heat flux---from the JR model are essentially the same as those for regular (elastic) fluids. 
Furthermore, the JR model was derived for nearly elastic ($e \approx 1$) granular fluids. 
Despite these limitations, the JR model has been widely exploited since its derivation and has been validated against particle simulations~\cite{BSSS1999} as well as against experiments~\cite{RBSS2001} for nearly elastic granular flows.
The JR model has also been employed in analyzing various instabilities in  granular flows. 
%
%
%
For instance,~\citet{ASL2008} investigated the shear-banding instability in two-dimensional dilute and dense granular shear flows using different variants of the JR model.  
In~\cite{ASL2008}, the authors essentially considered different models for the global equation of state (i.e.~for pressure) and for the shear viscosity (with and without a viscosity divergence term in its expression) while keeping the other transport coefficients the same as those in the original JR model, 
and they showed that some of their models underpin the shear banding while the others do not. Moreover, they also established that the onset of the shear-banding instabilities could be predicted with the knowledge of the pressure and shear viscosity of the system. This result indicates that 
the emergence of the shear banding is linked to the transport properties and, hence, to the choice of constitutive relations. Therefore, the selection of constitutive relations is crucial in describing instabilities and patterns in granular flows.

In another study on the the shear banding,~\citet{KM2006} substantiated that the experimentally observed shear-banding instabilities could not be perceived with the usual constitutive relations for dense granular gases; however, shear bands, miraculously, appear on slightly changing the  coefficient of the shear viscosity since the shear viscosity diverges at a lower density than the other transport properties. 
Note that the shear-banding instability predictions of both~\citet{ASL2008} and~\citet{KM2006} are valid only for the nearly elastic (or quasi-elastic) granular flows. 
However, in order to understand instability induced patterns in arbitrarily inelastic granular flows correctly, one must utilize proper constitutive relations which incorporate all the microscopic features of granular materials~\cite{Campbell1990,GoldI2003,Garzo2019}. 

To overcome the ``nearly elastic'' limitation of the JR model,~\citet{GD1999} performed the Chapman--Enskog expansion on the Enskog--Boltzmann equation and, in contrast to the JR model, obtained the restitution coefficient-dependent NSF-level constitutive relations for dense granular gases of hard spheres. 
Subsequently, the results of~\citet{GD1999} were generalized to an arbitrary dimension $d$ 
by~\citet{Lutsko2005}. We shall refer to the constitutive relations obtained in Refs.~\cite{GD1999} and \cite{Lutsko2005} by the GDL model.
The main differences between the JR and GDL models are as follows:  (i) the heat flux in the latter contains an additional term proportional to the density gradient that vanishes identically for elastic particles; this term is absent in the former and
(ii) all the transport coefficients depend on the restitution coefficient $e$ in the latter while only the collisional dissipation 
depends on $e$ in the former. As a consequence, the GDL model is not limited to nearly elastic granular fluids.
Recently,~\citet{ACSGP2013} conducted a comparative study of the Faraday instability in granular flows through the JR and GDL models and through the event-driven simulations, and also concluded that the choice of appropriate constitutive relations is crucial for analyzing granular patterns. 


There have been several theoretical studies on the shear-banding instability in a granular uniform shear flow (USF) for quasi-elastic particles, 
and thus all of them are valid only for nearly elastic granular fluids. 
Nonetheless, to the best of authors' knowledge, the shear-banding instability in arbitrarily inelastic dense granular flows has never been addressed theoretically. 
One of the reasons---among others---is that the granular NSF equations are valid strictly for processes involving small spatial gradients, and moreover since the (reduced) shear rate in the granular USF is inversely proportional to the collision frequency, small spatial gradients (the validity region of the NSF equations) would again mean large restitution coefficient (or nearly elastic particles) \cite{Lutsko2006, Garzo2006}.      
Furthermore, the granular USF state is inherently anisotropic and hence requires the \emph{generalized} transport coefficients for predicting the instabilities in the USF correctly~\cite{Garzo2006}. 
The generalized transport coefficients depend on the shear rate and have tensorial form. 
For dilute granular flows, the generalized transport coefficients were independently obtained by \citet{Lutsko2006} and \citet{Garzo2006}, and have been utilized by \citet{Garzo2006} to analyze the linear stability of the USF. 
From his linear stability analysis, \citet{Garzo2006} showed 
significant discrepancies between the results obtained with the
generalized transport coefficients and usual NSF transport coefficients; nonetheless, the unavailability of the numerical/experimental results thwarted  quantitative comparisons in~\cite{Garzo2006}. 
On the other hand, 
the derivation of such generalized transport coefficients in the case of dense granular flows is extremely involved and is beyond the scope of the present paper. 
Hence the NSF equations along with the GDL model~\cite{GD1999, Lutsko2005} for transport coefficients, which is the best known NSF-level hydrodynamic model for describing dense granular flows, can be considered as an intermediate way of investigating the linear stability of the USF.

In this context, the goal of the present work is to analyze the shear-banding instability in arbitrarily inelastic dense granular flows of hard disks by using the GDL constitutive relations. 
By means of the linear stability analysis of the USF, the onset of the shear-banding instability is predicted. 
In contrast to previous studies~\cite{KM2006,ASL2008},
the present stability results are valid for dilute-to-dense arbitrarily inelastic particles with the only assumption that the restitution coefficient is constant, i.e.~it does not depend on the impact velocity. 
It is emphasized that the focus of the present work is to determine the control parameters for the onset of the shear-banding instability but not the shape and location of a shear band, for which a theory, based on the principle of minimum energy dissipation, developed e.g.~in \cite{UTKW2004, TUKW2007} may be needed. 
%


The rest of the paper is organized as follows. The problem description and the governing equations are presented in Sec.~\ref{sec:Governing_Equations}. The non-dimensionalization and base state flow whose stability is to be investigated are demonstrated in Sec.~\ref{sec:non-dim_basestate}. The linear stability of the base state flow is analyzed in Sec.~\ref{sec:LSA}. The results and discussion are elucidated in Sec.~\ref{Sec:results}. The paper ends with conclusions and outlook in Sec.~\ref{sec:Conclusion}.


\section{Granular hydrodynamic equations}
\label{sec:Governing_Equations}
A granular flow of mono-disperse smooth identical inelastic 
hard disks 
of diameter $d_p$ can be described by 
the mass, momentum and energy balance equations, which 
read~\cite{BP2004,Garzo2019}
%
\begin{align}
\label{BalEqs}
\left.
\begin{aligned}
\left( \frac{\partial }{\partial \bar{t}} + \bar{\bm{U}} \cdot \bar{\bm{\nabla}} \right)\bar{\rho} 
+ \bar{\rho} \,\bar{\bm{\nabla}}\cdot \bar{\bm{U}} &= 0,
\\
\bar{\rho} \left( \frac{\partial }{\partial \bar{t}} +  \bar{\bm{U}} \cdot \bar{\bm{\nabla}} \right) \bar{\bm{U}} 
+ \bar{\bm{\nabla}}  \cdot \bar{\bm{P}}&=  \bm{0} ,
\\
\bar{\rho} \left( \frac{\partial }{\partial \bar{t}} + \bar{\bm{U}} \cdot \bar{\bm{\nabla}} \right) \bar{T} 
+ \bar{\bm{P}}:\bar{\bm{\nabla}} \bar{\bm{U}} 
+ \bar{\bm{\nabla}} \cdot \bar{\bm{q}}
&= - \bar{\mathcal{D}}. 
\end{aligned}
\right\}
\end{align}
Here, $\bar{\rho}=\rho_p \, \phi$ is the mass density with $\rho_p$ being the material density and $\phi$ being the volume fraction of grains; $\bar{\bm{U}}=(\bar{u},\bar{v},\bar{w})$ is the coarse-grained velocity with $\bar{u}$, $\bar{v}$ and $\bar{w}$ being its components in the $\bar{x}$-, $\bar{y}$- and $\bar{z}$-directions, respectively; $\bar{T}$ is the granular temperature;  
$\bar{\bm{P}}$ is the pressure tensor; $\bar{\bm{q}}$ is the (granular) heat flux; 
$\bar{\mathcal{D}}$ is the collisional dissipation due to inelastic collisions among grains; 
and $d$ denotes the dimension of the problem which takes value two for hard-disk flows and three for 
hard-sphere flows.


Clearly, system \eqref{BalEqs} is not closed due to the presence of the additional unknowns: $\bar{\bm{P}}$, $\bar{\bm{q}}$ and $\bar{\mathcal{D}}$. 
These unknowns are typically expressed in terms of the hydrodynamic variables $\bar{\rho}$, $\bar{\bm{U}}$, $\bar{T}$ and their spatial gradients by means of the Chapman--Enskog expansion, see e.g.~\cite{GS1995, BDKS1998, SG1998, GD1999, 
BP2004, Gupta2011}. 
To first order in spatial gradients of the hydrodynamic variables, these unknowns are expressed as~\cite{GS1995,SG1998}
\begin{align}
\label{PqD}
\bar{\bm{P}} &= \big( \bar{p} - \bar{\gamma} \, \bar{\bm{\nabla}} \cdot \bar{\bm{U}} \big) \, \bar{\bm{I}} - 2 \bar{\eta} \, \bar{\bm{S}} = \big( \bar{p} - \bar{\lambda} \, \bar{\bm{\nabla}} \cdot \bar{\bm{U}} \big) \, \bar{\bm{I}} - \bar{\eta} \big[\bar{\bm{\nabla}} \bar{\bm{U}} + (\bar{\bm{\nabla}} \bar{\bm{U}})^\mathsf{T}\big],
\\
\bar{\bm{q}} &= - \bar{\kappa} \, \bar{\bm{\nabla}} \bar{T} - \bar{\mu} \, \bar{\bm{\nabla}} \phi,
\\
\bar{\mathcal{D}} &= \bar{\mathcal{D}}_0 + \bar{\mathcal{D}}_1 \, \bar{\bm{\nabla}} \cdot \bar{\bm{U}},
\end{align}
where $\bar{\lambda} = \bar{\gamma} - 
\bar{\eta}$; $\bar{\bm{I}}$ is the identity tensor; and $\bar{\bm{S}}= \frac{1}{2} \big[\bar{\bm{\nabla}} \bar{\bm{U}} + (\bar{\bm{\nabla}} \bar{\bm{U}})^\mathsf{T}\big] - 
(\bar{\bm{\nabla}}\cdot \bar{\bm{U}}) \,\bar{\bm{I}}$ is the deviatoric strain rate tensor; the quantities $\bar{p}$, $\bar{\eta}$, $\bar{\gamma}$, $\bar{\kappa}$, $\bar{\mu}$, $\bar{\mathcal{D}}_0$ and $\bar{\mathcal{D}}_1$ are the pressure, shear viscosity, bulk viscosity, pseudo-thermal conductivity, Dufour-like coefficient (which identically vanishes for ordinary fluids), zeroth- and first-order contributions to the collisional dissipation, respectively, and are given in the form of constitutive relations: 
\begin{align}
\label{ConstRel}
\left.
\begin{aligned}
\bar{p}(\phi,\bar{T},e) &=  f_1(\phi,e)  \rho_p \bar{T},
\\ 
\bar{\eta}(\phi,\bar{T},e) &=  f_2(\phi,e) \, \rho_p d_p \sqrt{\bar{T}}, 
&\quad
\bar{\gamma}(\phi,\bar{T},e) &=  f_3(\phi,e) \, \rho_p d_p \sqrt{\bar{T}},
\\
\bar{\kappa}(\phi,\bar{T},e) &=  f_4(\phi,e) \, \rho_p d_p \sqrt{\bar{T}},
&\quad
\bar{\mu}(\phi,\bar{T},e) &=  f_{4h}(\phi,e) \, \rho_p d_p \bar{T}\sqrt{\bar{T}},
\\
\bar{\mathcal{D}}_0(\phi,\bar{T},e) &=  f_5(\phi,e) \, \frac{\rho_p}{d_p} \bar{T} \sqrt{\bar{T}},
&\quad
\bar{\mathcal{D}}_1(\phi,\bar{T},e) &=  f_{5u}(\phi,e) \, \rho_p \bar{T}.
\end{aligned}
\right\}
\end{align}
%
%
Here all the $f_i$'s are the dimensionless functions of the volume fraction $\phi$ and restitution coefficient $e$ only. It is worthwhile to note that the values of $f_i$'s are different for different models for the constitutive relations. 
From the GDL model, $f_i$'s in the case of hard-disk flows ($d=2$) read~\cite{Lutsko2005, Garzo2019}

%


\begin{align}
\label{finGDL}
\left.
\begin{aligned}
f_1(\phi,e) &= \phi \, \big[1+(1+e)\,G(\phi)\big],
\\
f_3(\phi,e) &= \frac{1}{\sqrt{\pi}} (1+e) \left(1 - \frac{c}{32}\right) \phi \, G(\phi),
\\
f_2(\phi,e) &= \frac{\sqrt{\pi}}{8} \frac{\left[1 - \frac{1}{4}(1+e)(1-3e)\,G(\phi)\right] \left[1 + \frac{1}{2}(1+e)\,G(\phi)\right]}{\nu_\eta^\ast - \frac{1}{2} \zeta_0^\ast} + \frac{1}{2} f_3(\phi),
\\
f_4(\phi,e) &= \frac{\sqrt{\pi}}{2} \left[1 + \frac{3}{4}(1+e)\,G(\phi)\right] \kappa_k^\ast + \frac{1}{\sqrt{\pi}} (1+e) \left(1+\frac{7c}{32}\right) \phi \, G(\phi),
\\
f_{4h}(\phi,e) &=  \frac{\sqrt{\pi}}{2\phi} \left[1 +\frac{3}{4}(1+e) \,G(\phi) \right] \mu_k^\ast, 
\\
f_5(\phi,e) &= \frac{4}{\sqrt{\pi}} (1-e^2) \left(1 + \frac{3 c}{32}\right) \phi \, G(\phi),
\\
f_{5u}(\phi,e) &= \frac{3}{2} (1-e^2)\left[\frac{3}{32} \frac{\frac{1}{8}\omega^\ast - \frac{c}{3}(1+e)(1-3e)}{\nu_\zeta^\ast - \frac{3}{4} (1-e^2)} - 1\right]  \phi \, G(\phi),
\end{aligned}
\right\}
\end{align} 
where $c$, the fourth cumulant estimating the lowest order correction to the Gaussian distribution function, is given by~\cite{vNE1998}
\begin{align}
c &= \frac{32(1-e)(1-2e^2)}{57-25e+30e^2-30e^3},
\end{align}
and $G(\phi) = \phi \, \chi(\phi)$ with $\chi(\phi)$ being the pair correlation function adopted from~\cite{Torquato1995}: 
\begin{equation}
\label{eqn:chi_T}
\chi(\phi) =
\begin{dcases}
 \frac{1 - 7 \phi/16}{(1-
 \phi)^2}& \mbox{for} \quad 0\leq \phi <\phi_f,
\\
 \frac{1 - 7 \phi_f/16}{(1-
 \phi_f)^2} 
 \left( \frac{\phi_c-\phi_f}{\phi_c-\phi}\right)
 & \mbox{for}\quad \phi_f\leq \phi \leq\phi_c.
\end{dcases}
\end{equation}
In \eqref{eqn:chi_T}, $\phi_f$ and $\phi_c$ are the freezing packing fraction and the random close-packing fraction, respectively. For hard-disk flows ($d=2$), $\phi_f \approx 0.69$ and $\phi_c \approx 0.82$ \cite{Torquato1995}.
The other variables in \eqref{finGDL} for hard-disk flows ($d=2$) are given by~\cite{Garzo2012}
\begin{align}
\label{astQuantities}
\left.
\begin{aligned}
\zeta_0^\ast &= \frac{1}{2} (1-e^2) \left(1+\frac{3c}{32}\right) \, \chi(\phi),
\\
\nu_\eta^\ast &= \frac{1}{8} (1+e) (7-3e) \left(1+\frac{7c}{32}\right) \, \chi(\phi),
\\
\kappa_k^\ast &= \frac{1+c + \frac{3}{8}(1+e)^2 \left[2e -1 + \frac{c}{2}(1+e)\right] G(\phi)}{2 \big(\nu_\kappa^\ast -2 \zeta_0^\ast \big)},
\\
\mu_k^\ast &= \frac{\zeta_0^\ast \kappa_k^\ast (1 + \phi \, \partial_\phi{\ln{\chi}}) + \frac{c}{4} + \frac{3}{8}(1+e) \left(1 + \frac{1}{2} \phi \, \partial_\phi{\ln{\chi}}\right)  \left[e(e -1) + \frac{c}{12}(14-3e+3e^2)\right] G(\phi)}{\nu_\kappa^\ast - \frac{3}{2} \zeta_0^\ast},\\
\nu_\kappa^\ast &= \frac{1}{4} (1+e) \left[1 + \frac{15}{4} (1-e) + \frac{365-273e}{128}c\right] \, \chi(\phi),
\\
\omega^\ast &= (1+e) \left[(1-e^2) (5e-1) + \frac{c}{12}(41 - 69 e + 3 e^2 - 15 e^3)\right],
\\
\nu_\zeta^\ast &= \frac{1}{192} (1+e)(185 - 153 e + 30 e^2 - 30 e^3).
\end{aligned}
\right\}
\end{align}
The quantities in~\eqref{astQuantities} emanate from the so-called modified Sonine approximation introduced by~\citet{GSM2007}, and are presented in a more coherent form---for an arbitrary dimension $d$---by~\citet{Garzo2012} (see also the textbook~\cite{Garzo2019}).
All the quantities, except $\mu_k^\ast$, in~\eqref{astQuantities} are also given in~\cite{ACSGP2013} for hard-disk flows ($d=2$). 
The $\mu_k^\ast$ in the present work is twice of that of~ \cite{ACSGP2013} but is the same as that in~\cite{Garzo2012} for $d=2$ 
in order to keep the standard form of the reduced Dufour-like coefficient ($\mu^\ast = n \mu/(\kappa_0 T)$) given, e.g., in~\cite{BDKS1998, Garzo2012, GST2018, Garzo2019}. 



Notably, the physical properties in a granular flow are transported via two mechanisms, namely the kinetic and collisional. 
The former is attributed to streaming, i.e.~to the movement of particles from one place to another, while the latter to collisions among grains.
Needless to say, the former is dominant in dilute flows while the latter in dense flows. 
Owing to these transport mechanisms,
the pressure tensor $\bar{\bm{P}}$ and heat flux $\bar{\bm{q}}$ for a (dense) granular flow can be decomposed into their kinetic and collisional contributions, i.e.~$\bar{\bm{P}} = \bar{\bm{P}}^{k} + \bar{\bm{P}}^{c}$ and $\bar{\bm{q}} = \bar{\bm{q}}^{k} + \bar{\bm{q}}^{c}$, where the superscripts `$k$' and `$c$' denote the kinetic and collisional contributions, respectively, see e.g.~\cite{GD1999, Lutsko2005, GSM2007} and references therein. 
Accordingly, the pressure $\bar{p}$ and the transport coefficients $\bar{\eta}, \bar{\gamma}, \bar{\kappa}, \bar{\mu}$ and, hence the dimensionless functions $f_i$'s for $i\in\{1,2,3,4,4h\}$, can be decomposed into their kinetic and collisional parts, i.e.~$f_i = f_i^k + f_i^c$. 
From the expressions of $\bar{\bm{P}}^{k}$, $\bar{\bm{P}}^{c}$, $\bar{\bm{q}}^{k}$ and $\bar{\bm{q}}^{c}$ given in~\cite{Lutsko2005, Garzo2012}, it is straightforward to determine the kinetic and collisional parts of $f_i$'s, which read
\begin{align}
\label{finGDL_0}
\left.
\begin{aligned}
f_1^k(\phi,e) &= \phi,
\\
f_3^k(\phi,e) &= 0,
\\
f_2^k(\phi,e) &= \frac{\sqrt{\pi}}{8} \frac{1 - \frac{1}{4}(1+e)(1-3e)\,G(\phi)}{\left(\nu_\eta^\ast - \frac{1}{2} \zeta_0^\ast\right)},
\\
f_4^k(\phi,e) &= \frac{\sqrt{\pi}}{2}\kappa_k^\ast 
\\
f_{4h}^k(\phi,e) &= \frac{\sqrt{\pi} }{2 \phi}  \mu_k^\ast, 
\end{aligned}
\right\}
\end{align} 
and
\begin{align}
\label{finGDL_d}
\left.
\begin{aligned}
f_1^c(\phi,e) &= (1+e)\,\phi \, G(\phi),
\\
f_3^c(\phi,e) &= \frac{1}{\sqrt{\pi}} (1+e) \left(1 - \frac{c}{32}\right) \phi \, G(\phi),
\\
f_2^c(\phi,e) &= \frac{1}{2}(1+e)\,G(\phi) f_2^k +
\frac{1}{2} f_3(\phi),
\\
f_4^c(\phi,e) &= \frac{3\sqrt{\pi}}{8}(1+e)\,G(\phi) 
\kappa_k^\ast 
+ \frac{1}{\sqrt{\pi}} (1+e) \left(1+\frac{7c}{32}\right) \phi \, G(\phi),
\\
f_{4h}^c(\phi,e) &=  \frac{3 \sqrt{\pi}}{8} (1+e) \, \chi(\phi)  \mu_k^\ast.
\end{aligned}
\right\}
\end{align} 
Indeed, as expected, the collisional contributions to the pressure tensor ($\bar{\bm{P}}^c$) and heat flux ($\bar{\bm{q}}^c$) for dilute granular flows ($\phi \to 0$) vanish and hence the pressure tensor and heat flux for dilute granular flows are given by $\bar{\bm{P}} = \bar{\bm{P}}^{k}$ and $\bar{\bm{q}} = \bar{\bm{q}}^{k}$~\cite{Garzo2013, Garzo2019}. Consequently, $f_i^k$'s in~\eqref{finGDL_0} can be referred to as the dilute limit of $f_i$'s for $i\in\{1,2,3,4,4h\}$.
The collisional dissipation $\bar{\mathcal{D}}$ in a (dilute or dense) granular flow, on the other hand, is attributed only to inelastic collisions among grains. Interestingly, the first-order contribution (in spatial gradients) to the collisional dissipation, $\bar{\mathcal{D}}_1$, is zero for dilute granular flows~\cite{Garzo2013, Garzo2019}, and the zeroth-order contribution (in spatial gradients) to the collisional dissipation, $\bar{\mathcal{D}}_0$, for dilute granular flows can be obtained from its dense counterpart given, e.g., in~\cite{Lutsko2005, Garzo2012, Garzo2013, Garzo2019} by taking $\chi(\phi \to 0) = 1$. Consequently, the dimensionless functions $f_5$ and $f_{5u}$ for dilute granular hard-disk flow ($d=2$) are given by
\begin{align}
\label{f5dilute}
f_5(\phi \to 0,e) = \frac{4}{\sqrt{\pi}} (1-e^2) \left(1 + \frac{3 c}{32}\right) \phi^2
\quad\textrm{and}\quad
f_{5u}(\phi \to 0,e)=0.
\end{align}
In what follows, the constitutive relations~\eqref{ConstRel} obtained using $f_i \approx f_i^k$
for $i\in\{1,2,3,4,4h\}$ with $f_i^k$ from~\eqref{finGDL_0} and $f_5, f_{5u}$ from~\eqref{f5dilute}
will be referred to as the dilute limit of the GDL model.
Similarly, the constitutive relations~\eqref{ConstRel} obtained using $f_i \approx f_i^c$ for $i\in\{1,2,3,4,4h\}$ with $f_i^c$ from~\eqref{finGDL_d} and $f_5, f_{5u}$ from~\eqref{finGDL} will be referred to as the collisional limit of the GDL model. 

\section{Non-dimensionalization and base state}
\label{sec:non-dim_basestate}

\begin{figure}[!ht]
\includegraphics[scale=1]
{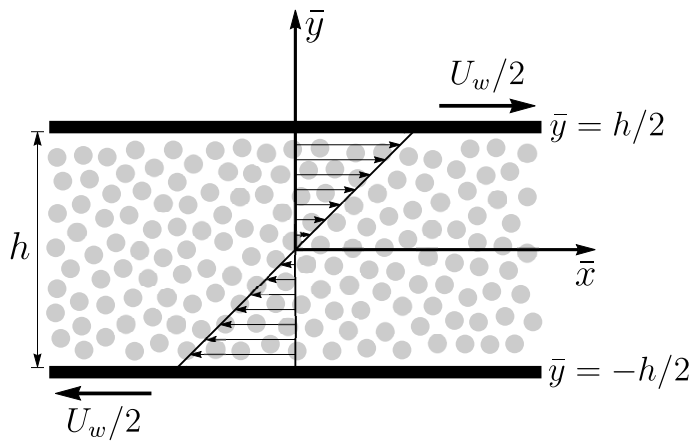}
\caption{Schematic of the uniform shear flow of granular hard disks confined in a channel.  
}
\label{fig:schematic}
\end{figure}

We shall investigate a plane shear flow of granular hard disks 
confined in a two-dimensional channel of width $h$. 
The flow is driven by the two oppositely moving walls (of the channel) placed at $\bar{y}=\pm h/2$ with a speed $U_w/2$ along the streamwise ($\bar{x}$) direction; 
hence the overall shear rate is $U_w/h$, see figure~\ref{fig:schematic}.

It is worthwhile to note that the velocity slip at a rigid wall is inherently present in granular flows. This has also been verified through experiments and computer simulations \cite{Mueth2000, CG2004, CLG2006}. 
The velocity slip, in turn, leads to the generation of the pseudo-thermal energy at the wall that competes with the energy lost due to particle-wall collisions. 
Consequently, energy flux at the wall is, in general, non zero.
Nevertheless, for simplicity, we choose the no-slip and zero heat flux (adiabatic) boundary conditions for the velocity and temperature, respectively.
These idealized boundary conditions have also been used in previous studies, e.g.~in \cite{WJS1996, AN1998, Alam2006, ASL2008,Khain2007}, pertaining to the stability of granular shear flows. 
The no-slip and adiabatic boundary conditions for the problem under consideration read 
\begin{align}
\label{BCswithdimensions}
\bar{u}\left(\pm \frac{h}{2}\right) = \pm \frac{U_w}{2}
\quad \textrm{and} \quad
\left. \frac{\partial \bar{T}}{\partial \bar{y}} \right|_{\bar{y}=\pm h/2} = 0.
\end{align}

For the purpose of non-dimensionalization, 
we choose $h$ as a reference length, 
$U_w$ as a reference velocity, the inverse of the shear rate $h/U_w$ as a reference time and $\rho_p$ as the reference mass density. 
In the following, the quantities with overbar are dimensional and their bare counterparts are dimensionless.
%
Let us define the dimensionless quantities without bars as follows,
\begin{align}
\bm{\nabla} = h \bar{\bm{\nabla}}, \quad \bm{U} = \frac{1}{U_w} \bar{\bm{U}}, \quad t = \frac{U_w}{h} \bar{t}, \quad \rho=\frac{\bar{\rho}}{\rho_p}, \quad T = \frac{H^2}{ U_w^2} \bar{T},
\end{align}
where $H=h/d$ is referred to as the dimensionless gap between the walls or the channel width. 
With $\rho_p$ as the scaling for the density, the volume fraction 
$\phi$ also denotes dimensionless density.    
With these scales, governing equations~\eqref{BalEqs} along with the constitutive relations~\eqref{ConstRel} in the dimensionless form read
\begin{align}
\label{dimlessBalEqs}
\left.
\begin{aligned}
\left( \frac{\partial }{\partial t} + \bm{U} \cdot \bm{\nabla} \right)\phi  &= -\phi  \, \bm{\nabla} \cdot \bm{U},
\\
\phi \left( \frac{\partial }{\partial t} +  \bm{U} \cdot \bm{\nabla} \right) \bm{U} &= \frac{1}{H^2} \bigg[ -  \bm{\nabla} p + \bm{\nabla}  \left(  \lambda \, \bm{\nabla} \cdot \bm{U} \right) + \bm{\nabla}  \cdot \Big\{ \eta \, \Big(\bm{\nabla}  \bm{U} + (\bm{\nabla}  \bm{U})^\mathsf{T}\Big) \Big\}  \bigg],
\\
\frac{d}{2} \, \phi \left( \frac{\partial }{\partial t} + \bm{U} \cdot \bm{\nabla} \right) T &=  
\frac{1}{H^2} \bm{\nabla} \cdot  \left(\kappa \, \bm{\nabla} T + \mu \, \bm{\nabla} \phi\right) -  p \, \bm{\nabla} \cdot \bm{U} + \lambda \, (\bm{\nabla} \cdot \bm{U})^2 
\\
&\quad + \eta \, \Big\{\bm{\nabla}  \bm{U} + (\bm{\nabla}  \bm{U})^\mathsf{T}\Big\}: \bm{\nabla} \bm{U} 
- \mathcal{D}_0 - \mathcal{D}_1  \, \bm{\nabla} \cdot \bm{U},
\end{aligned}
\right\}
\end{align}
where
\begin{align}
\label{dimlessTransCoeff}
\left.
\begin{aligned}
p(\phi,T,e) &= f_1(\phi,e) T,    &  \eta(\phi,T,e) &= f_2(\phi,e)\, \sqrt{T}, 
\\
\gamma(\phi,T,e) &= f_3(\phi,e)\, \sqrt{T}, & \lambda(\phi,T,e) &= \gamma - \frac{2}{d}\eta, 
\\
\kappa(\phi,T,e) &= f_4(\phi,e)\, \sqrt{T}, & 
\mu(\phi,T,e) &=  f_{4h}(\phi,e)\, T \sqrt{T}, 
\\
\mathcal{D}_0(\phi,T,e) &=  f_5(\phi,e)\, T \sqrt{T},
\qquad & 
\mathcal{D}_1(\phi,T,e) &= f_{5u}(\phi,e)\, T.
\end{aligned}
\right\}
\end{align}
Boundary conditions~\eqref{BCswithdimensions} in the dimensionless form read 
\begin{align}
\label{BCseothoutdimensions}
{u}\left(\pm \frac{1}{2}\right) = \pm\frac{1}{2}
\quad \mbox{and}
\quad
\left. \frac{\partial {T}}{\partial {y}} \right|_{{y}=\pm 1/2} = 0.
\end{align}


We shall be investigating the 
shear banding in a two-dimensional plane shear flow.
Specifically, we focus on the gradient banding due to which bands of dense and dilute regions form along the gradient (i.e.~$y$-) direction. 
Such instability arises from the perturbations having no variation in the streamwise (i.e.~$x$-) direction. 
The streamwise independent (i.e. $\partial (\cdot)/\partial x = 0$) governing equations~\eqref{dimlessBalEqs} in the dimensionless form read
\begin{align}
\label{2deqs}
\left.
\begin{aligned}
\frac{\partial \phi}{\partial t} + v \frac{\partial \phi}{\partial y} + \phi \frac{\partial v}{\partial y} &= 0,
\\
\phi \left( \frac{\partial u}{\partial t} + v \frac{\partial u}{\partial y} \right) &= \frac{1}{H^2} \frac{\partial}{\partial y} \left( \eta \frac{\partial u}{\partial y} \right),
\\
\phi \left( \frac{\partial v}{\partial t} + v \frac{\partial v}{\partial y} \right) &= \frac{1}{H^2} \left[ -\frac{\partial p}{\partial y} + \frac{\partial}{\partial y} \left( \lambda \frac{\partial v}{\partial y} \right) + 2 \frac{\partial}{\partial y} \left( \eta \frac{\partial v}{\partial y} \right) \right],
\\
\phi \left( \frac{\partial T}{\partial t} + v \frac{\partial T}{\partial y} \right) &= \frac{1}{H^2} \frac{\partial}{\partial y} \left( \kappa \frac{\partial T}{\partial y} + \mu \frac{\partial \phi}{\partial y} \right) - p \frac{\partial v}{\partial y} + \lambda \left(\frac{\partial v}{\partial y}\right)^2 + \eta \left(\frac{\partial u}{\partial y}\right)^2 
+ 2 \eta \left(\frac{\partial v}{\partial y}\right)^2
- \mathcal{D}_0 - \mathcal{D}_1 \frac{\partial v}{\partial y}.
\end{aligned}
\right\}
\end{align}

\subsection{Base state: Uniform shear flow}
\label{subsec:basestate}

The basic flow, whose stability is to be analyzed here, is assumed to be a steady $(\partial(\cdot)/\partial t \equiv 0)$, fully developed $(\partial(\cdot)/\partial x \equiv 0)$, plane shear flow 
of
the following form
\begin{align}
\phi = \phi(y),\quad u = u(y),\quad v=0,\quad T=T(y).
\end{align}
%
Hence, the mass balance equation~\eqref{2deqs}$_1$ is identically satisfied for this flow, while the remaining equations in \eqref{2deqs} reduce to
\begin{align}
\frac{\mathrm{d}}{\mathrm{d} y} \left( \eta \frac{\mathrm{d}  u}{\mathrm{d} y} \right) =0,
\quad
 -\frac{\mathrm{d}  p}{\mathrm{d} y}=0,
\quad
\frac{1}{H^2} \frac{\mathrm{d}}{\mathrm{d} y} \left( \kappa \frac{\mathrm{d}  T}{\mathrm{d} y} + \mu \frac{\mathrm{d} \phi}{\mathrm{d} y} \right) + \eta \left(\frac{\mathrm{d}  u}{\mathrm{d} y}\right)^2 
- \mathcal{D}_0=0.
\end{align}
Using~\eqref{dimlessTransCoeff}, the above equations, respectively, lead to
\begin{align}
\left.
\begin{gathered}
f_2(\phi,e) \sqrt{ T} \, \frac{\partial  u}{\partial y}=c_1,
\qquad
f_1(\phi,e)\, T = c_2,
\\
\frac{1}{H^2} \frac{\mathrm{d}}{\mathrm{d} y} \left( f_4(\phi,e) \sqrt{ T} \, \frac{\mathrm{d}  T}{\mathrm{d} y} + f_{4h}(\phi,e) \,  T\sqrt{ T} \frac{\mathrm{d} \phi}{\mathrm{d} y} \right) + f_2(\phi,e) \sqrt{ T} \left(\frac{\mathrm{d}  u}{\mathrm{d} y}\right)^2 
- f_5(\phi,e) \,  T\sqrt{ T} =0,
\end{gathered}
\right\}
\end{align}
%
%
where $c_1$ and $c_2$ are the integration constants.
The above set of equations with no-slip and zero heat flux boundary conditions~\eqref{BCseothoutdimensions} admits the 
following base state solution:
\begin{align}
\label{baseState}
\phi^0(y)= \phi_0,
\quad
u^0(y) = y,
\quad 
v^0(y) = 0, 
\quad
T^0(y) = f_2^0 / f_5^0 = T_0,
\end{align}
where the superscript `0' 
denotes the base state solutions, 
$\phi_0$ and $T_0$ are constants, and $f_2^0=f_2(\phi^0,e)$ and $f_5^0=f_5(\phi^0,e)$. 
Note that the base flow velocity is linear with constant density and constant temperature. Such a base state gives constant or uniform shear rate, i.e.~$u^0_y=1$, thus
leading to the USF. The base state density $\phi_0$, the channel width $H$ and the restitution coefficient $e$ are the control parameters for the problem. 
\section{Linear stability analysis}
\label{sec:LSA}
%



%
%
\subsection{Perturbation equations}
For the linear stability analysis of the USF, the field variables $(\phi,u,v,T)$ are decomposed into the base state solution plus the perturbation from the base state solution as follows:
\begin{align}
\label{pert}
\begin{aligned}
\phi(t,y) = \phi^0(y) + \phi^\prime(t,y),
\quad
u(t,y) = u^0 (y)+ u^\prime(t,y),
\quad
v(t,y) = v^0(y) + v^\prime(t,y),
\quad
T(t,y) = T^0(y) + T^\prime(t,y).
\end{aligned}
\end{align}
Here, the field variables with prime denote the perturbations from their respective base states. 
These perturbations are assumed to be small so that the linear theory remains valid. 
Substituting 
the field variables from~\eqref{pert} 
into governing equations~\eqref{2deqs} and retaining only the linear terms of the perturbed field variables, one obtains the governing equations for the perturbed field variables:
\begin{align}
\left.
\begin{aligned}
\frac{\partial \phi^\prime}{\partial t} &= - \phi_0 \frac{\partial v^\prime}{\partial y},
\\
\frac{\partial u^\prime}{\partial t} &=  \frac{\eta_{\phi}^0}{\phi_0 H^2} \frac{\partial \phi^\prime}{\partial y} + \frac{\eta^0}{\phi_0 H^2} \frac{\partial^2 u^\prime}{\partial y^2} - v^\prime + \frac{\eta_T^0}{\phi_0 H^2} \frac{\partial T^\prime }{\partial y},
\\
\frac{\partial v^\prime}{\partial t} &=   -\frac{p_{\phi}^0}{\phi_0  H^2}   \frac{\partial \phi^\prime}{\partial y} +  \frac{\lambda^0 + 2 \eta^0}{\phi_0  H^2}  \frac{\partial^2 v^\prime}{\partial y^2}  - \frac{p_T^0}{\phi_0  H^2} \frac{\partial T^\prime}{\partial y},
\\
\frac{\partial T^\prime}{\partial t}
&= \frac{1}{\phi_0}\left(\frac{\mu^0}{H^2}  \frac{\partial^2}{\partial y^2} + \eta_{\phi}^0 - \mathcal{D}_{0,\phi}^0\right) \phi^\prime 
+ \frac{2 \eta^0}{\phi_0} \frac{\partial u^\prime}{\partial y} - \frac{p^0 + \mathcal{D}_1^0}{\phi_0} \frac{\partial v^\prime}{\partial y} + \frac{1}{\phi_0}\left(\frac{\kappa^0}{H^2}  \frac{\partial^2}{\partial y^2} + \eta_T^0 - \mathcal{D}_{0,T}^0\right) T^\prime,
\end{aligned}
\right\}
\end{align}
where the subscripts `$\phi$' and `$T$' represent the partial derivatives 
with respect to $\phi$ and $T$, respectively,  
and the superscript zero represents the variables calculated at the base state.
The above equations can be written as a matrix system 
\begin{align}
\label{eigProb}
\frac{\partial \bm{X}}{\partial t} = \mathcal{L} \, \bm{X},
\end{align}
where $\bm{X} = (\phi^\prime,u^\prime,v^\prime,T^\prime)^\mathsf{T}$ is the vector of perturbed fields and $\mathcal{L}$ is the matrix of linear differential 
operators given by
\begingroup
\renewcommand{\arraystretch}{3}
\begin{align}
\label{Loperator}
\mathcal{L} =
\begin{bmatrix}
0 & 0 & - \phi_0 \dfrac{\partial}{\partial y} & 0
\\
\dfrac{\eta_{\phi}^0}{\phi_0 H^2} \dfrac{\partial}{\partial y} & \dfrac{\eta^0}{\phi_0 H^2} \dfrac{\partial^2}{\partial y^2} & - 1 & \dfrac{\eta_T^0}{\phi_0 H^2} \dfrac{\partial}{\partial y}
\\
-\dfrac{p_{\phi}^0}{\phi_0  H^2} \dfrac{\partial }{\partial y} & 0 & \dfrac{\lambda^0 + 2 \eta^0}{\phi_0  H^2}  \dfrac{\partial^2}{\partial y^2}  & - \dfrac{p_T^0}{\phi_0  H^2} \dfrac{\partial}{\partial y}
\\
\dfrac{1}{\phi_0}\left(\dfrac{\mu^0}{H^2}  \dfrac{\partial^2}{\partial y^2} + \eta_{\phi}^0 - \mathcal{D}_{0,\phi}^0\right) & 
\dfrac{2 \eta^0}{\phi_0} \dfrac{\partial}{\partial y} & - \dfrac{p^0 + \mathcal{D}_1^0}{\phi_0} \dfrac{\partial}{\partial y} & \dfrac{1}{\phi_0}\left(\dfrac{\kappa^0}{H^2}  \dfrac{\partial^2}{\partial y^2} + \eta_T^0 - \mathcal{D}_{0,T}^0\right)
\end{bmatrix}.
\end{align}
\endgroup
From~\eqref{BCseothoutdimensions}, the boundary conditions for the perturbed field variables read
\begin{equation}
\label{BCsPerturbed}
u^\prime\left(\pm \frac{1}{2}\right) =\left. \frac{\partial T^\prime}{\partial y} \right|_{y=\pm 1/2} = 0. 
%
\end{equation}

\subsection{Analytical solutions}

We perform the standard linear stability analysis on the USF by assuming a normal mode solution of the form
$\bm{X}(t,y)=\hat{\bm X}(y) \exp(\omega t)$, where $\hat{\bm X}=(\hat{\phi},\hat{u},\hat{v},\hat{T})^\mathsf{T}$ and
$\omega=\omega_r + \mathsf{i}\, \omega_i$ is the complex frequency whose real part $\omega_r$ represents growth or decay rate of the perturbations and imaginary part $\omega_i$ denotes the oscillation of the perturbation, with $\mathsf{i}$ being the imaginary unit.  
Substituting this normal mode solution into the 
linearized perturbation equations~\eqref{eigProb}--\eqref{BCsPerturbed}, we get the following  matrix eigenvalue problem:
\begin{equation}
\hat{L} \hat{\bm X} = \omega  \hat{\bm X},\qquad \hat{u}\left(\pm \frac{1}{2}\right)=\frac{{\rm d}\hat{T}}{{\rm d}y}=0,
\label{matevp}
\end{equation}
where 
$\hat{L}  = \mathcal{L} \left(
\partial/\partial y \rightarrow
{\rm d}/{\rm d}y, \partial^2/\partial y^2 \rightarrow {\rm d}^2/{\rm d}y^2
\right)
$. 
It has been verified that eigenvalue problem~\eqref{matevp} has an analytical solution in terms of sine and cosine 
functions~\cite{AN1998}, as
\begin{align}
\label{analyticalSol}
(\hat{\phi},\hat{T})^\mathsf{T} &= (\phi_1,T_1)^\mathsf{T} \cos{[\pi\,\beta \, (y \pm 1/2)]},
\quad
(\hat{u},\hat{v})^\mathsf{T} = (u_1,v_1)^\mathsf{T} \sin{[\pi\,\beta \, (y \pm 1/2)]},
\end{align}
where $\beta = 1,2,3,\dots$ are the mode numbers and $(\phi_1, u_1, v_1, T_1)^\mathsf{T}$ is the constant amplitude of the normal mode solution.  
For instance, 
$\beta=1$ is the fundamental mode and
$\beta=2$ is the second harmonic of the normal mode solution, etc. 
With solution~\eqref{analyticalSol}, problem~\eqref{matevp}$_1$ simplifies to
\begin{align}
\label{simplifiedEigVal}
L_1 \bm{X}_1 =\omega \bm{X}_1 ,
\end{align}
where 
$\bm{X}_1 =({\phi}_1,{u}_1,{v}_1,{T}_1)^\mathsf{T}$ and
\begingroup
\renewcommand{\arraystretch}{3}
\begin{align}
\label{Qmat}
L_1 = 
\begin{bmatrix}
0 & 0 & - \pi\,\beta\,\phi_0 & 0
\\
-\dfrac{\pi\,\beta\,\eta_{\phi}^0}{\phi_0 H^2} & -\dfrac{\pi^2\beta^2 \eta^0}{\phi_0 H^2} & - 1 & -\dfrac{\pi\,\beta\,\eta_T^0}{\phi_0 H^2} 
\\
\dfrac{\pi\,\beta\,p_{\phi}^0}{\phi_0  H^2} & 0 & -\dfrac{\pi^2\beta^2 (\lambda^0 + 2 \eta^0)}{\phi_0  H^2} & \dfrac{\pi\,\beta\,p_T^0}{\phi_0  H^2} 
\\
\dfrac{1}{\phi_0} \left(-\dfrac{\pi^2\beta^2 \mu^0}{H^2} + \eta_{\phi}^0 - \mathcal{D}_{0,\phi}^0\right) & 
\dfrac{2 \pi\,\beta\,\eta^0}{\phi_0} & - \dfrac{\pi\,\beta\,(p^0 + \mathcal{D}_1^0)}{\phi_0} & \dfrac{1}{\phi_0}\left(-\dfrac{\pi^2\beta^2\kappa^0}{H^2} + \eta_T^0 - \mathcal{D}_{0,T}^0\right)
\end{bmatrix}.
\end{align}
\endgroup
For the nontrivial solutions of~\eqref{simplifiedEigVal} 
${\rm det}(L_1 - \omega I)=0$, where $I$ is the identity matrix of size 4. This condition is the dispersion relation and
can be written as
\begin{align}
\label{DISPREL}
\omega^4 +a_3 \omega^3 +a_2 \omega^2+a_1 \omega + a_0=0,
\end{align}
where
\begin{align}
\label{ai}
\begin{aligned}
a_0 = \frac{1}{H^4} a_{04} + \frac{1}{H^6} a_{06},
\quad
a_1 = \frac{1}{H^2} a_{12} + \frac{1}{H^4} a_{14} + \frac{1}{H^6} a_{16},
\quad
a_2 = \frac{1}{H^2} a_{22} + \frac{1}{H^4} a_{24}, 
\quad 
a_3 = a_{30} + \frac{1}{H^2} a_{32}.
\end{aligned}
\end{align}
Here the coefficients $a_{ij}$ are the functions of the transport coefficients evaluated at the base state. However, their explicit expressions 
are relegated to appendix~\ref{dis_coeff} for better readability. 
Dispersion relation~\eqref{DISPREL} is a fourth-degree polynomial in $\omega$ with real coefficients and, therefore, there are three possibilities for four roots of~\eqref{DISPREL}: (i) all roots are real, (ii) two complex conjugate pairs of roots and (iii) two real roots and one complex conjugate pair of roots.

\subsection{Asymptotic analysis}
\label{subsec:asy_ana}
\begin{figure}[!htb]
\includegraphics[width=0.42\textwidth]
{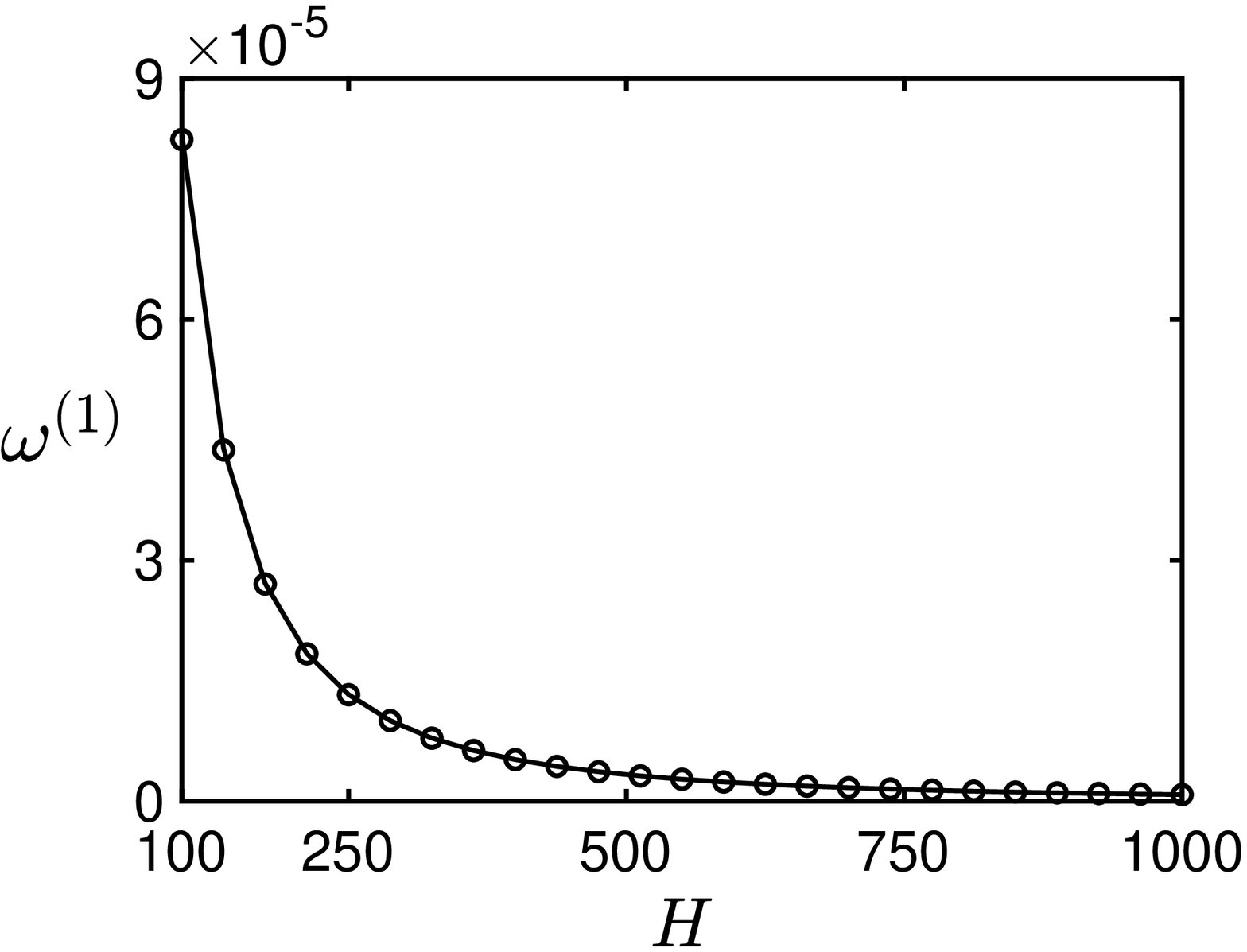}
\quad
\includegraphics[width=0.42\textwidth]
{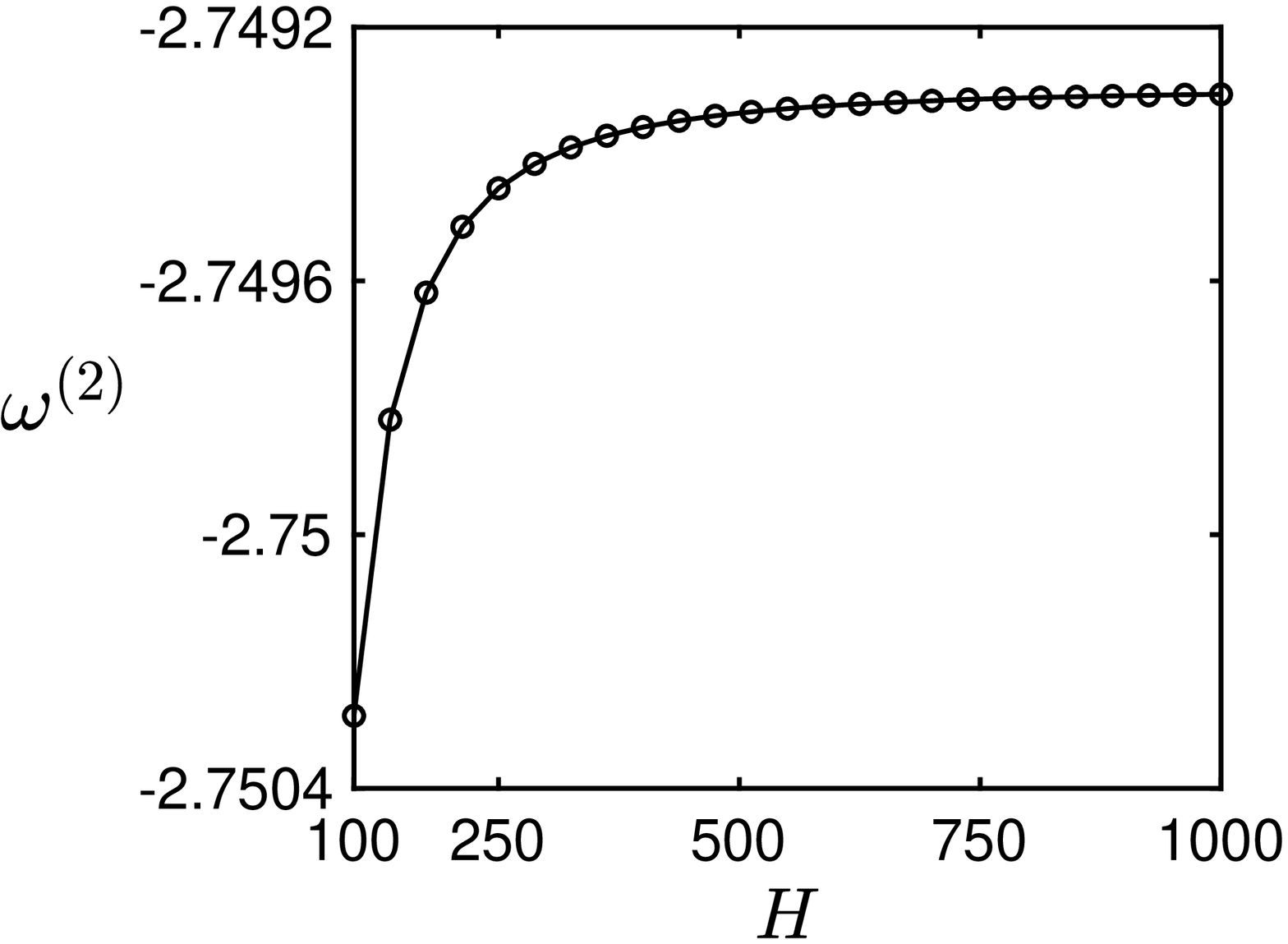}
\quad
\\[2ex]
\includegraphics[width=0.42\textwidth]
{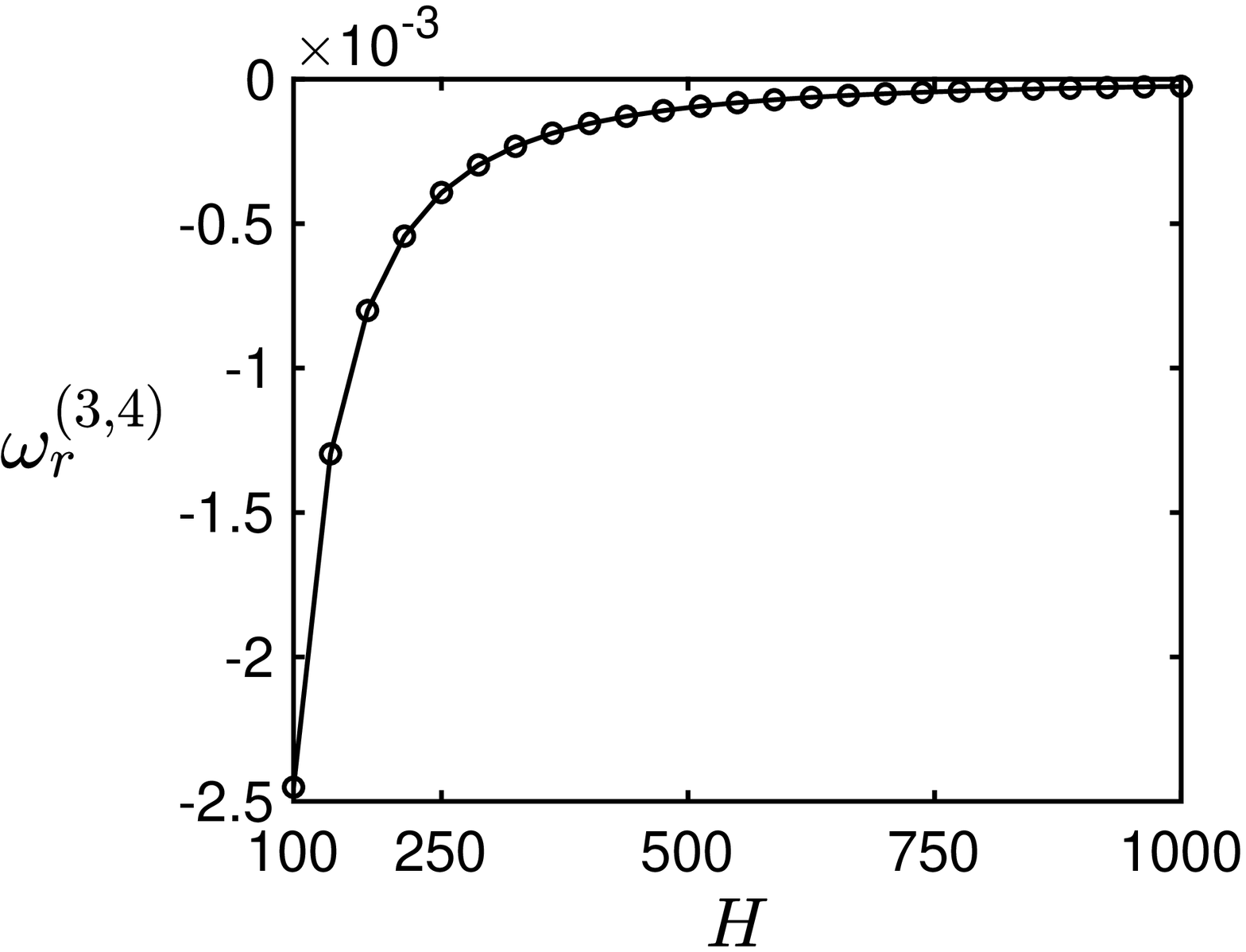}
\qquad
\includegraphics[width=0.42\textwidth]
{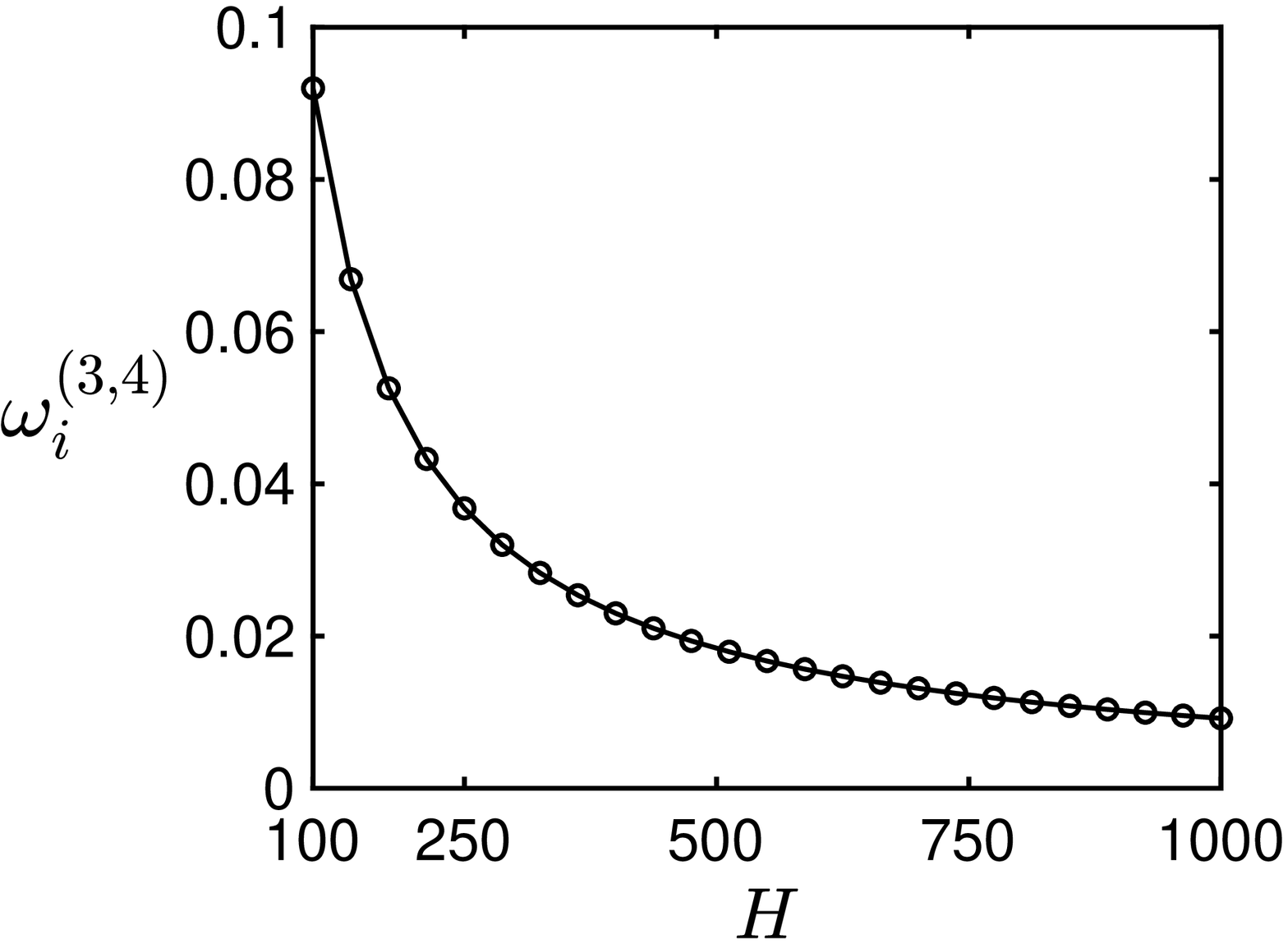}
\caption{The eigenvalues of \eqref{simplifiedEigVal} for large $H$ for parameters $\phi^0=0.6$, $e=0.5$ and $\beta=1$. The solid lines denote the eigenvalues from \eqref{omega1}--\eqref{omega4}, which were obtained through the asymptotic analysis of \eqref{DISPREL}, and the symbols delineate those obtained by solving matrix eigenvalue problem~\eqref{simplifiedEigVal} numerically. }
\label{fig:asymptotic_analysis}
\end{figure}

With the help of the classical asymptotic analysis in powers of
$H^{-1}$ with $H^{-1}\to 0$, one can find the analytical expressions of the eigenvalues (the roots of \eqref{DISPREL}), as follows.  
%
%
%
Let the frequency $\omega$ be represented by 
\begin{align}
\omega = \omega_0 + \frac{1}{H}\omega_1 +  \frac{1}{H^2}\omega_2 +  \frac{1}{H^3}\omega_3 + \dots,
\label{eqn:omega_exp}
\end{align}
where $\omega_0, \omega_1, \omega_2,\dots$ are unknown coefficients. 
Substituting this ansatz along with~\eqref{ai} into \eqref{DISPREL}, and comparing each power of $H$ on both sides of the resulting equations, one obtains algebraic equations, which are solved for the unknowns $\omega_i$'s in~\eqref{eqn:omega_exp}.  
Exploiting these values, 
one obtains four roots $\omega=\omega^{(1,2,3,4)}$ from~\eqref{DISPREL}, which are given by
\begin{align}
\omega^{(1)} &= - \frac{1}{H^2} \frac{a_{04}}{a_{12}} + \mathcal{O}(H^{-4}),
\label{omega1}
\\
\omega^{(2)} &= -a_{30} - \frac{1}{H^2} \frac{a_{12} - a_{22}a_{30} + a_{32}a_{30}^2}{a_{30}^2} + \mathcal{O}(H^{-4}),
\\
\omega^{(3,4)} &=  \omega_r^{(3,4)} \pm \mathsf{i} \, \omega_i^{(3,4)}
+ \mathcal{O}(H^{-4}),
\label{omega4}
\end{align}
where the subscripts `$r$' and `$i$' 
represent the real and imaginary parts, respectively, of the roots and
\begin{align}
\omega_r^{(3,4)} &= \frac{1}{H^2} \frac{a_{04} + (a_{12}^2/a_{30}^2) - (a_{12}a_{22}/a_{30})}{2a_{12}},
\label{omega_23}
\\
\omega_i^{(3,4)} &=
\sqrt{\frac{a_{12}}{a_{30}}} \left[ \frac{1}{H} - \frac{1}{H^3}
\left\{  \frac{1}{2} \frac{a_{32}}{a_{30}}
- \frac{3}{4} \frac{a_{22}}{a_{30}^2}
+\frac{5}{8} \frac{a_{12}}{a_{30}^3}
+\frac{1}{8} \frac{2a_{04}+a_{22}^2}{a_{12}a_{30}}
- \frac{1}{2} \frac{a_{14}}{a_{12}}
+\frac{1}{4} \frac{a_{04}a_{22}}{a_{12}^2}
-\frac{3}{8} \frac{a_{04}^2 a_{30}}{a_{12}^3}
\right\}
\right]
.
\end{align}
It is evident from~\eqref{omega1}--\eqref{omega4} that in the limit of large $H$, dispersion relation~\eqref{DISPREL} has two real roots $\omega^{(1,2)}$ and a complex conjugate pair of roots $\omega^{(3,4)}$. 
In the limit of large $H$, it is verified numerically that $\omega^{(2)}$ and $\omega_r^{(3,4)}$ are always negative resulting into the least stable mode as $\omega^{(1)}$. 
Figure~\ref{fig:asymptotic_analysis} illustrates the four eigenvalues for large $H$ obtained through the asymptotic analysis of dispersion relation~\eqref{DISPREL} (solid line) and those obtained by solving~\eqref{simplifiedEigVal} numerically (symbols) for $\phi^0=0.6$, $e=0.5$ and $\beta=1$. 
It can be seen from the figure that the eigenvalues from both the methods are in excellent agreement for large $H$.

\section{Results}
\label{Sec:results}

The linear stability of the USF of granular materials has been studied 
previously for the case of nearly elastic particles~\cite{AN1998,Alam2005a,
Alam2005,Alam2006,ASL2008}.
In this section, we analyze the linear stability of the USF of arbitrary inelastic granular hard disks
by appropriately choosing the GDL constitutive relations pertaining to dense granular flows~\cite{GD1999,Lutsko2005}.
In order to perform the linear stability analysis, eigenvalue problem~\eqref{matevp} is solved 
numerically using the Chebyshev spectral collocation method. 
In addition, eigenvalue problem~\eqref{simplifiedEigVal}, which is obtained using the  
%
%
exact solutions~\eqref{analyticalSol}, is also solved analytically.  
It is verified that the eigenvalues obtained 
numerically using the Chebyshev spectral collocation method and analytically using the exact solutions~\eqref{analyticalSol} are found in an excellent agreement (figure is not shown for brevity). 
Therefore in the present analysis, the eigenvalues are computed by solving~\eqref{simplifiedEigVal}. 
%
All the computations have been performed in \textsc{Matlab}\textsuperscript{\textregistered}. For analyzing the stability results, we define the least stable eigenvalue $\omega^l$ as one of the eigenvalues whose real part is maximum for a fixed mode number, and the dominant eigenvalue $\omega^d$ is one of the least stable eigenvalues whose real part is maximum over all the mode numbers $\beta$, i.e.
\begin{align}
\omega^l := \underset{\omega_r}\max\,\, \omega
\quad \textrm{and} \quad
\omega^d := \underset{ \beta}\max\,\, \omega^l.
\end{align}

\begin{figure}[!t]
\includegraphics[height = 67mm]
{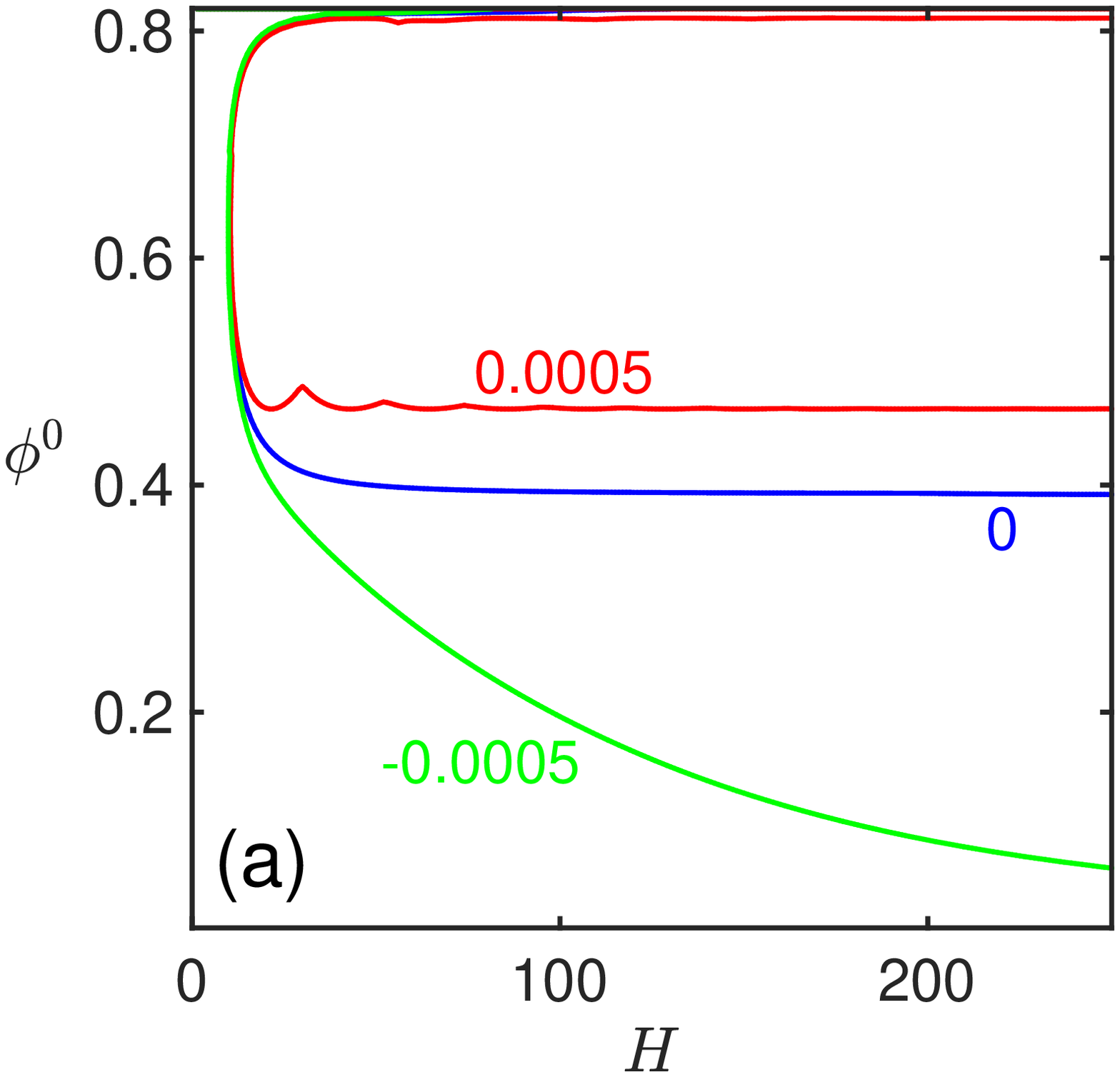}
\qquad\qquad\qquad
\includegraphics[scale=0.5]
{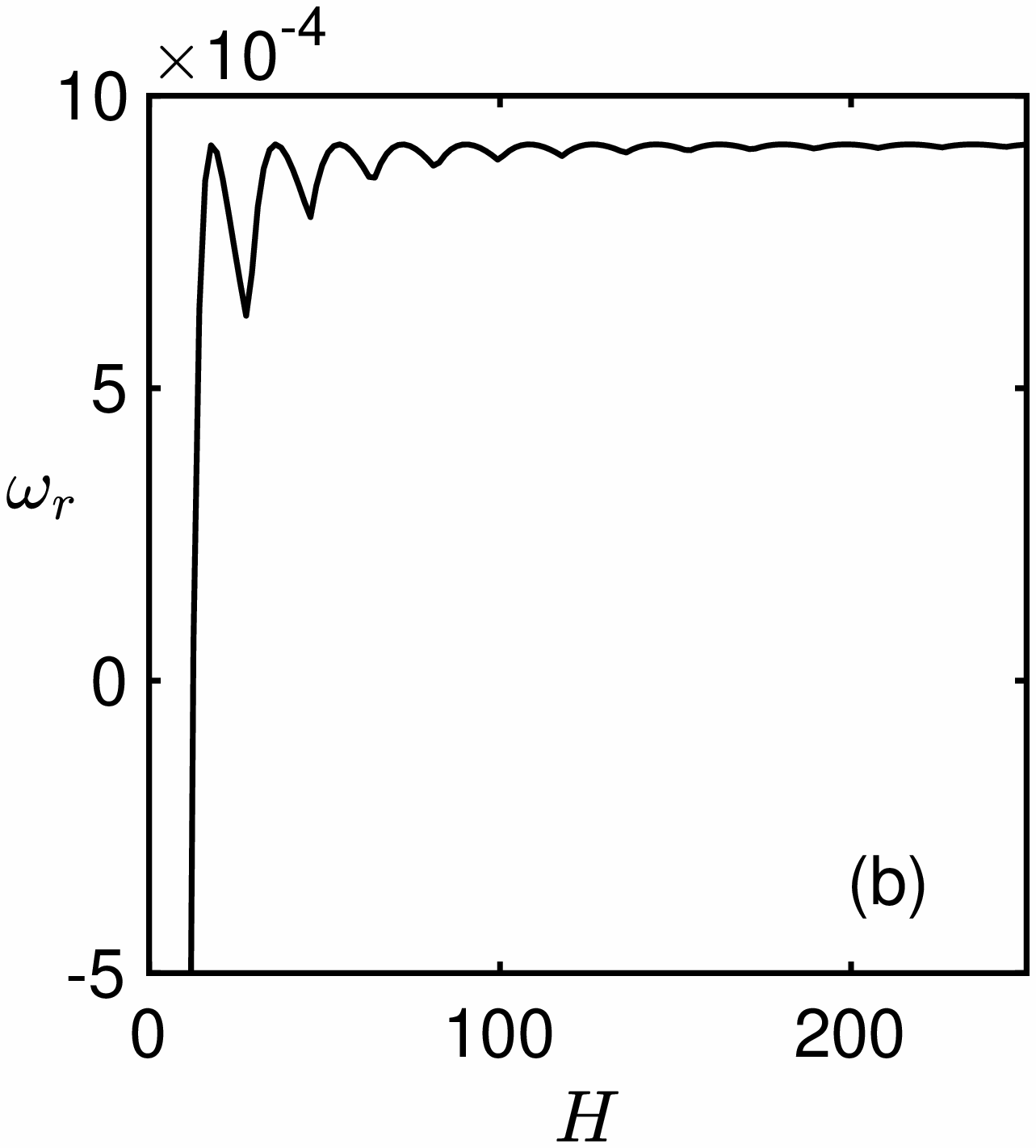}
\caption{(a) Contours of the positive, negative and zero growth rates in the $(H,\phi^0)$-plane, 
and (b) the variation of the growth rate with the channel width $H$ for $\phi^0=0.5$. 
The restitution coefficient is set to $e=0.6$. 
}
\label{fig1:contours_ep5}
\end{figure}

In order to avoid the mode number and grid dependencies on the stability predictions, the dominant modes are analyzed in the present work.
%
The contours of positive, negative and zero dominant growth rates, $\omega^d_r$, are shown in the $(H, \phi^0)$-plane for the inelastic particles with the restitution coefficient $e=0.6$, see figure~\ref{fig1:contours_ep5}(a). 
The real part of the dominant mode is positive (i.e.~$\omega^d_r>0$) inside the zero contour and negative (i.e.~$\omega^d_r<0$) outside the zero contour, 
therefore the flow is unstable inside the zero contour and stable outside.
%
It is also verified that the instability depicted in figure~\ref{fig1:contours_ep5}(a) is due to the stationary waves since the imaginary part of the complex frequency $\omega^d$ is always zero. 
This can also be seen from the asymptotic analysis presented in ~Sec.~\ref{subsec:asy_ana} that the least stable eigenvalue $\omega^l$ is always real. Therefore the shear-banding instabilities in a granular shear flow are due to the stationary waves. 

We have also examined that the lowest mode number, i.e.~$\beta=1$, is the first one to become unstable for a fixed restitution coefficient. 
This is similar to the classical Rayleigh--B{\'e}nard convection in which the first mode is the dominant mode~\cite{Chandrasekhar1961}. 
Figure~\ref{fig1:contours_ep5}(a) also illustrates that for a fixed restitution coefficient there exist a critical channel width $H_c(\phi^0)$ and a critical mean density $\phi^0_c(H)$ above which the USF is unstable and below which it is stable. 
That is to say, the USF becomes unstable if either $H>H_c(\phi^0)$ or $
\phi^0>\phi_c^0(H)$. 
In addition, it can also be seen from figure~\ref{fig1:contours_ep5}(a) that the USF remains stable for all densities and for all channel widths below an onset value of the mean density, say $\phi^0_{onset}$ (see Sec.~\ref{subsec:onset} for more details).  
Note that the onset mean density $\phi^0_{onset}$ depends only on the restitution coefficient.  
In particular, for $e=0.6$, the flow is stable when
$\phi^0<\phi^0_{onset} \approx 0.38$ for all values of $H$. 
Figure~\ref{fig1:contours_ep5}(a) also reveals that the USF is stable in the dilute limit ($\phi^0 \to 0$).

In order to get more insight, the variation of the dominant growth rate with the channel width for fixed values of the mean density and restitution coefficient is illustrated in 
figure~\ref{fig1:contours_ep5}(b).
The kinks (crests) in the figure correspond to the eigenmode crossing from the mode number $j$ to $j+1$, where $j=1,2,3,\dots$ is a positive integer. 
It is seen that the dominant mode number increases with the channel width, i.e.~$\beta=1$ mode is the first one to lose stability and remains unstable with increasing channel width 
until it crosses $\beta=2$ mode (first kink in figure~\ref{fig1:contours_ep5}(b)), thereafter $\beta=2$ mode becomes dominant mode until it crosses the next $\beta=3$ mode (second kink in figure~\ref{fig1:contours_ep5}(b)), and so on.  
For parameter values shown in figure~\ref{fig1:contours_ep5}(b), 
the USF becomes unstable for $H>H_c\approx 10.446$.




\begin{figure}[!htbp]
\includegraphics[scale=0.43]
{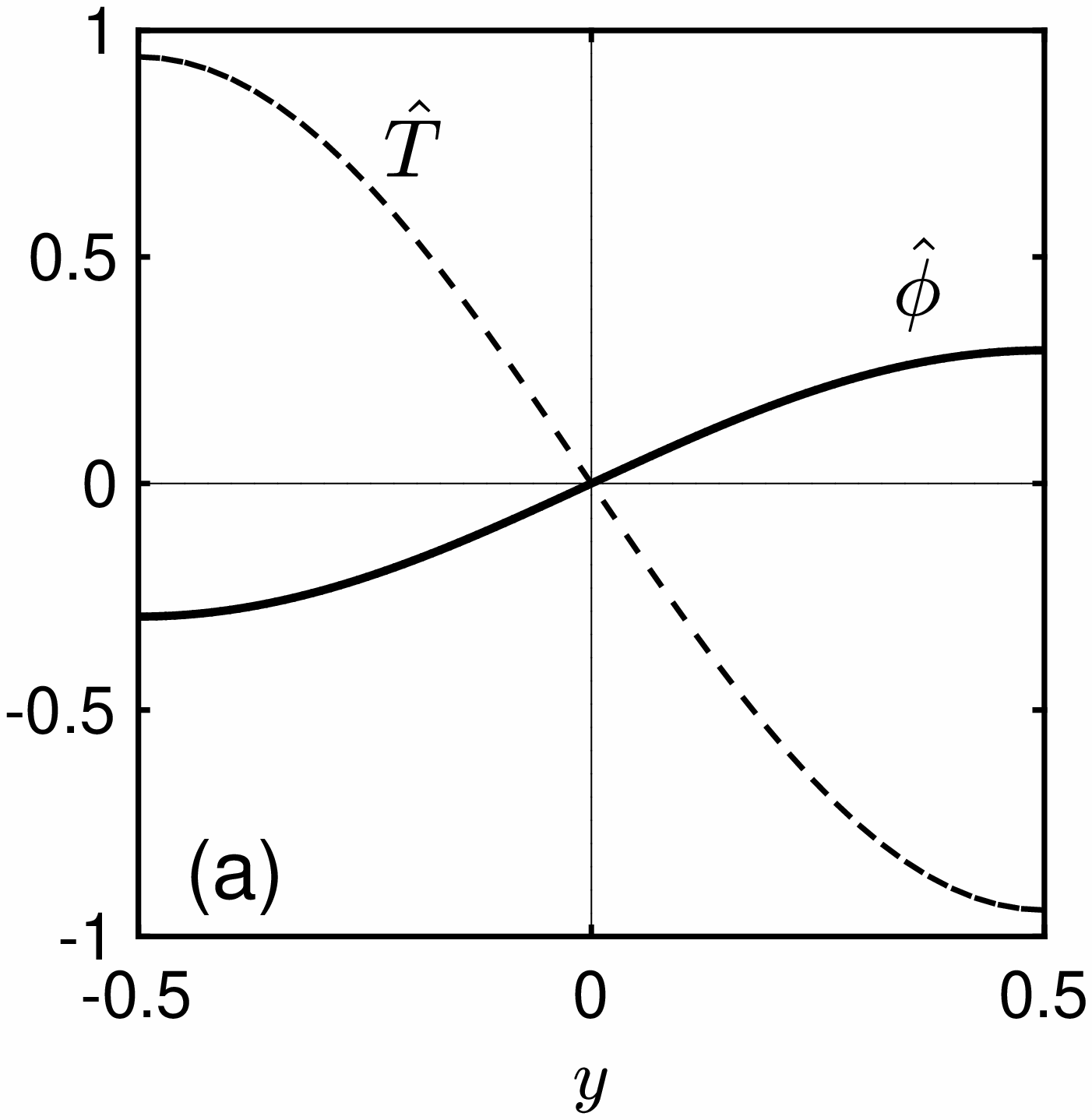}\qquad\quad\quad
\includegraphics[scale=0.43]
{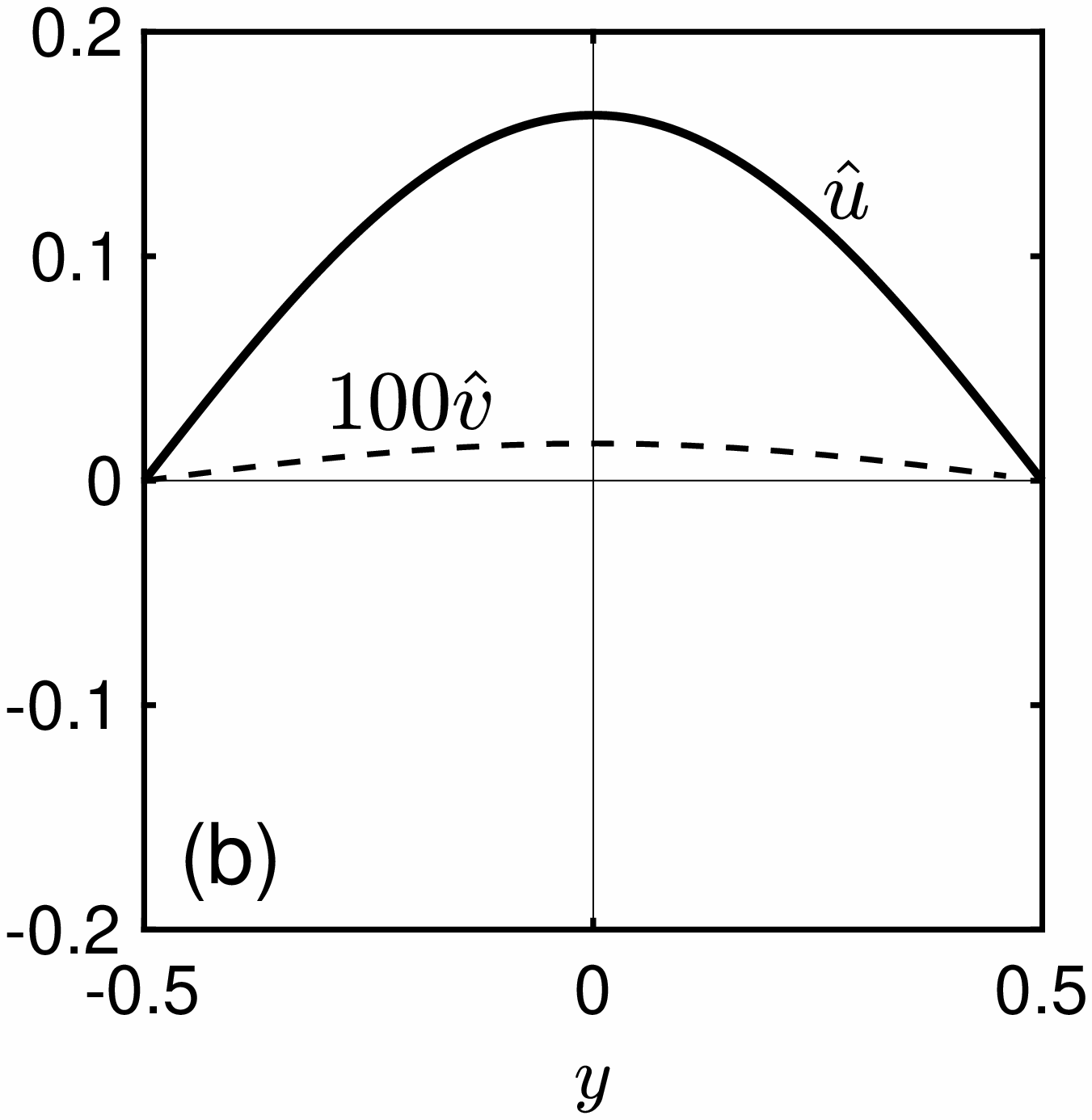}
\\[1em]
\includegraphics[scale=0.43]
{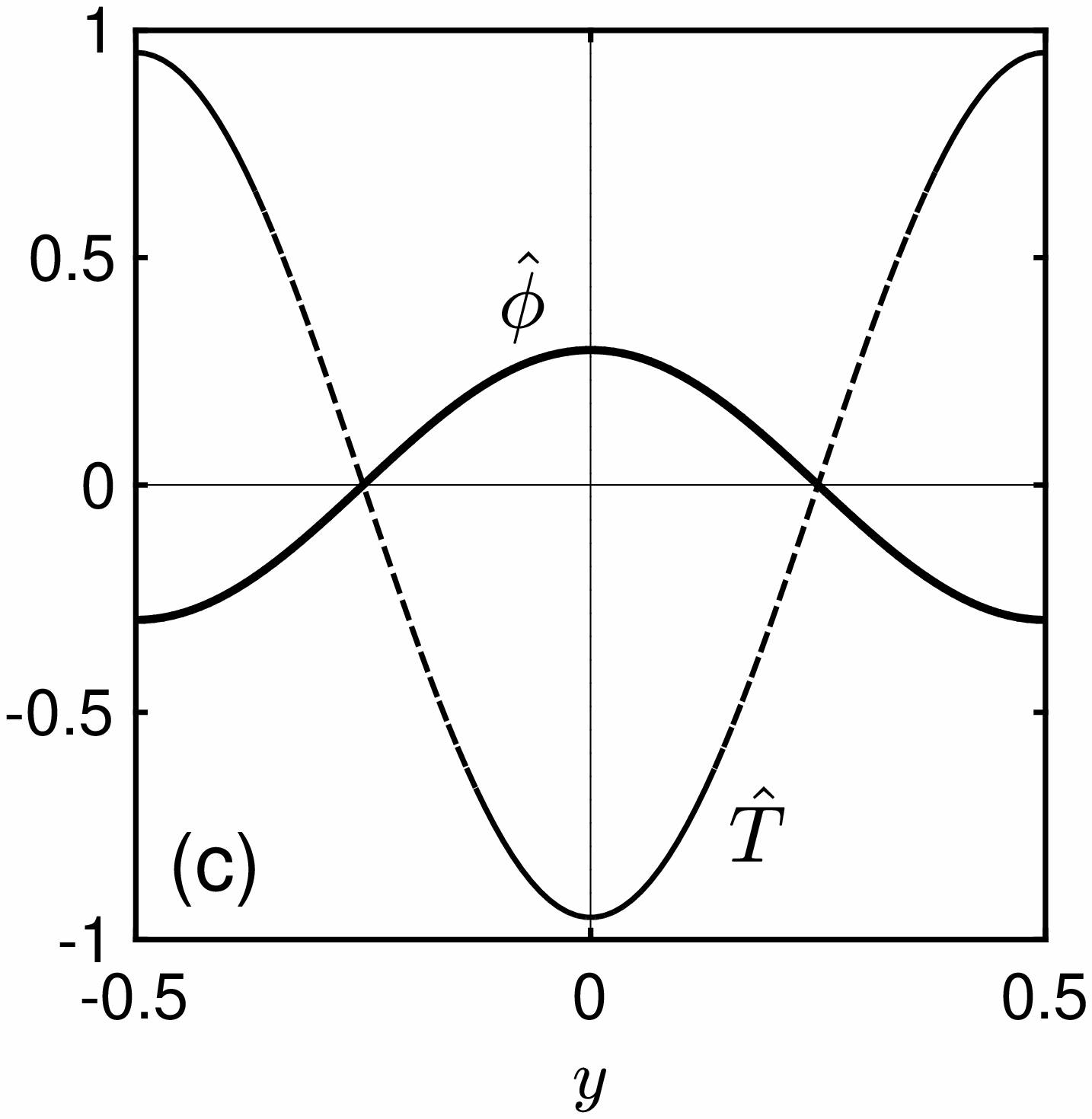}\qquad\quad\quad
\includegraphics[scale=0.43]
{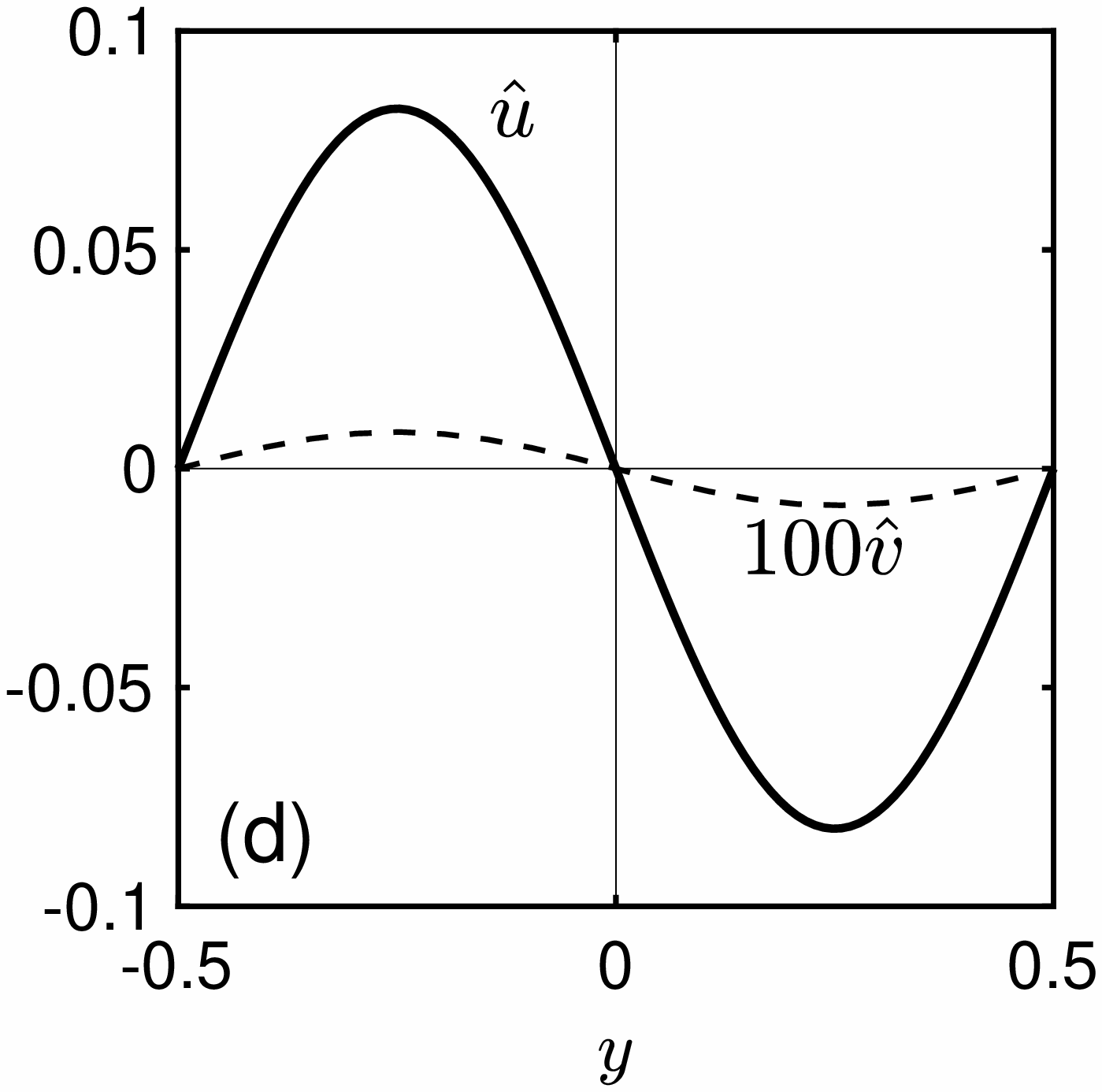}
\\[1em]
\includegraphics[scale=0.43]
{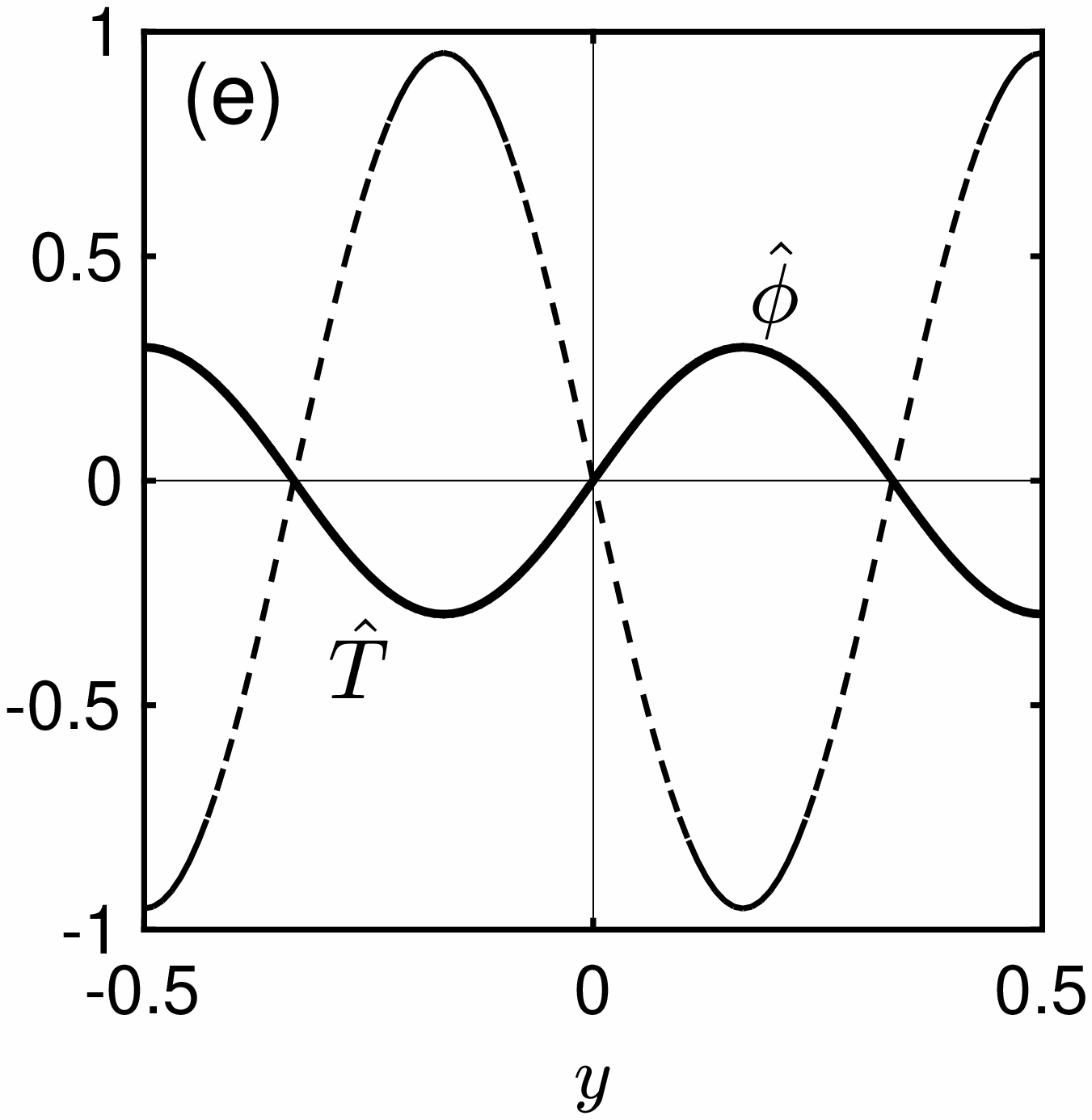}\qquad\quad\quad
\includegraphics[scale=0.43]
{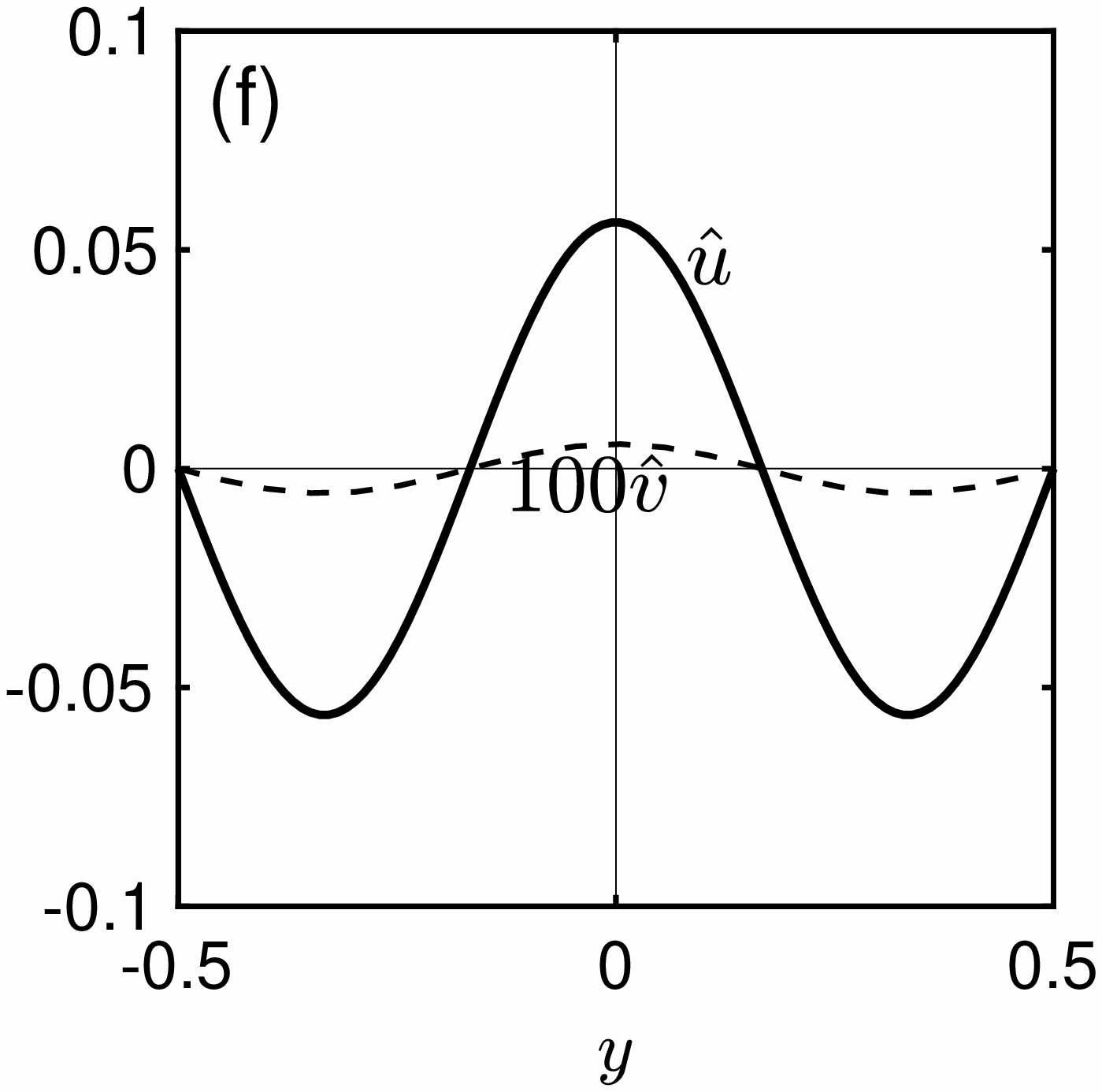}
\caption{Eigenfunctions~\eqref{analyticalSol} for three values of the channel width $H=$ 
$20$ (first row), $40$ (second row) and $50$ (third row) corresponding to $\beta=1$, $2$ and $3$, respectively. The other parameter values are the same as those in figure~\ref{fig1:contours_ep5}(b).}
\label{fig:solutions}
\end{figure}

Figure~\ref{fig:solutions} illustrates the density, temperature and velocity eigenfunctions for three values of the channel width, $H=20$, $40$, $50$, with other parameters being the same as those in figure~\ref{fig1:contours_ep5}(b). 
The eigenfunctions displayed in the first row correspond to $\beta =1$ mode for which $(\hat{\phi}, \hat{T}) \propto \cos(\pi y\pm\pi/2)=\sin(\pi y)$ 
that vanish once at $y=0$ in the flow domain, 
therefore the density and temperature solutions attain their mean values $\phi^0$ and $T^0$ once in the flow domain. 
The second row of figure~\ref{fig:solutions} corresponds to $\beta=2$ mode which gives eigenfunctions $(\hat{\phi}, \hat{T}) \propto \cos(2 y\pi\pm\pi)=\cos(2 \pi y)$ that vanish twice along the flow domain at $y=\pm 1/4$. Similarly, the third row corresponds to $\beta=3$ mode having the eigenfunctions $(\hat{\phi}, \hat{T}) \propto \cos(3 y\pi\pm3 \pi/2)=\pm \sin(3 \pi y)$ that vanish thrice at $y=0,~\pm 1/3$. 
Consequently, the density and temperature solutions attain its mean values once, twice and thrice in the flow domain for $H=20$, $40$ and $60$, respectively. 
By analyzing the corresponding velocity components (second column in figure~\ref{fig:solutions}), it is evident that while the horizontal component of the velocity eigenfunction $\hat{u}$ varies significantly, its transverse component $\hat{v}$ varies only slightly.

\begin{figure}[!t]
\centering{\includegraphics[scale=0.42]{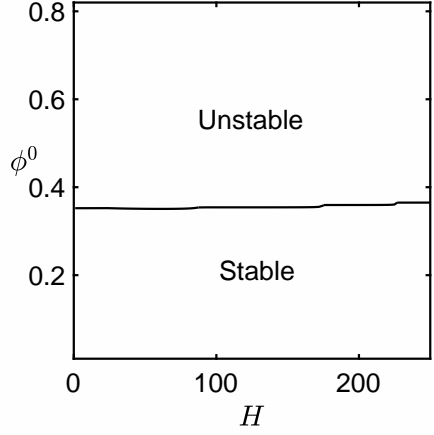}}
\caption{Neutral stability curve in the $(H,\phi^0)$-plane 
obtained with the dilute limit of the GDL model (i.e.~by using $f_i \approx f_i^k$ for $i\in\{1,2,3,4,4h\}$ with $f_i^k$ from \eqref{finGDL_0} and $f_5, f_{5u}$ from \eqref{f5dilute} in the constitutive relations~\eqref{ConstRel}).
The restitution coefficient is $e=0.6$.
\label{fig_contour_kinetic}
}
\end{figure}

Figure~\ref{fig_contour_kinetic} exhibits the neutral stability curve obtained with the dilute limit of the GDL model (i.e.~by using $f_i \approx f_i^k$ for $i\in\{1,2,3,4,4h\}$ with $f_i^k$ from \eqref{finGDL_0} and $f_5, f_{5u}$ from \eqref{f5dilute} in the constitutive relations~\eqref{ConstRel}). 
It is evident from the figure that the USF is always stable in the dilute limit ($\phi^0 \to 0$), which is also known from the previous studies~\cite{ASL2008,CG2004,CLG2006}. 
Therefore the dilute limit of the GDL model is able to capture the stability of dilute granular shear flows.
On the other hand, although the collisional mechanism is dominant over the kinetic mechanism in dense granular flows, the collisional limit of the GDL model (the constitutive relations~\eqref{ConstRel} obtained using $f_i \approx f_i^c$ for $i\in\{1,2,3,4,4h\}$ with $f_i^c$ from \eqref{finGDL_d} and $f_5, f_{5u}$ from \eqref{finGDL}) alone fails to capture the shear-banding instability in dense granular shear flows~\cite{AL2003a, CG2004, CLG2006} correctly  
(phase diagram is not shown for brevity): the collisional limit of the GDL model predicts that the USF is stable for high densities whereas the full GDL model (i.e.~the constitutive relations with \eqref{finGDL}) (see figure~\ref{fig1:contours_ep5}) and molecular dynamics simulations~\cite{AL2003a, CG2004, CLG2006} predict the shear-banding instability for dense granular shear flows. 
By analyzing three variants of the GDL model viz.~the full model, its dilute limit and its collisional limit, 
one can conclude that the choice of constitutive relations plays an important role in determining the shear-banding instability. For correct instability predictions in dense granular flows, both the kinetic and collisional mechanisms are important and hence none of them should be discarded. 
Therefore, in the following, we focus on the effect of the inelasticity on the shear-banding instability through the full GDL model.

\begin{figure}[!ht]
\begin{minipage}[T]{0.4\textwidth}
\includegraphics[scale=0.42]{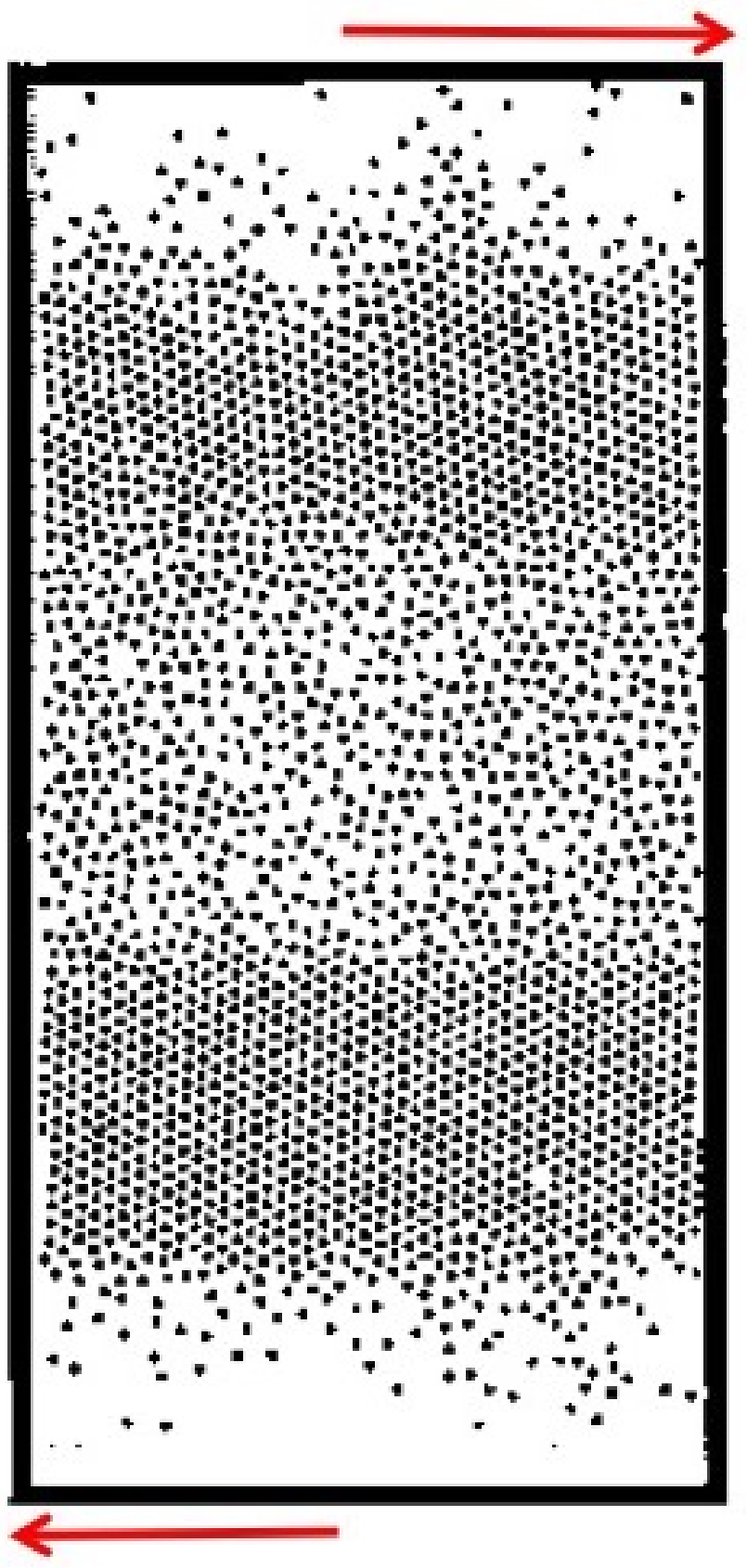}
\vspace*{7mm}
\end{minipage}
\begin{minipage}[B]{0.45\textwidth}
\includegraphics[scale=0.57]{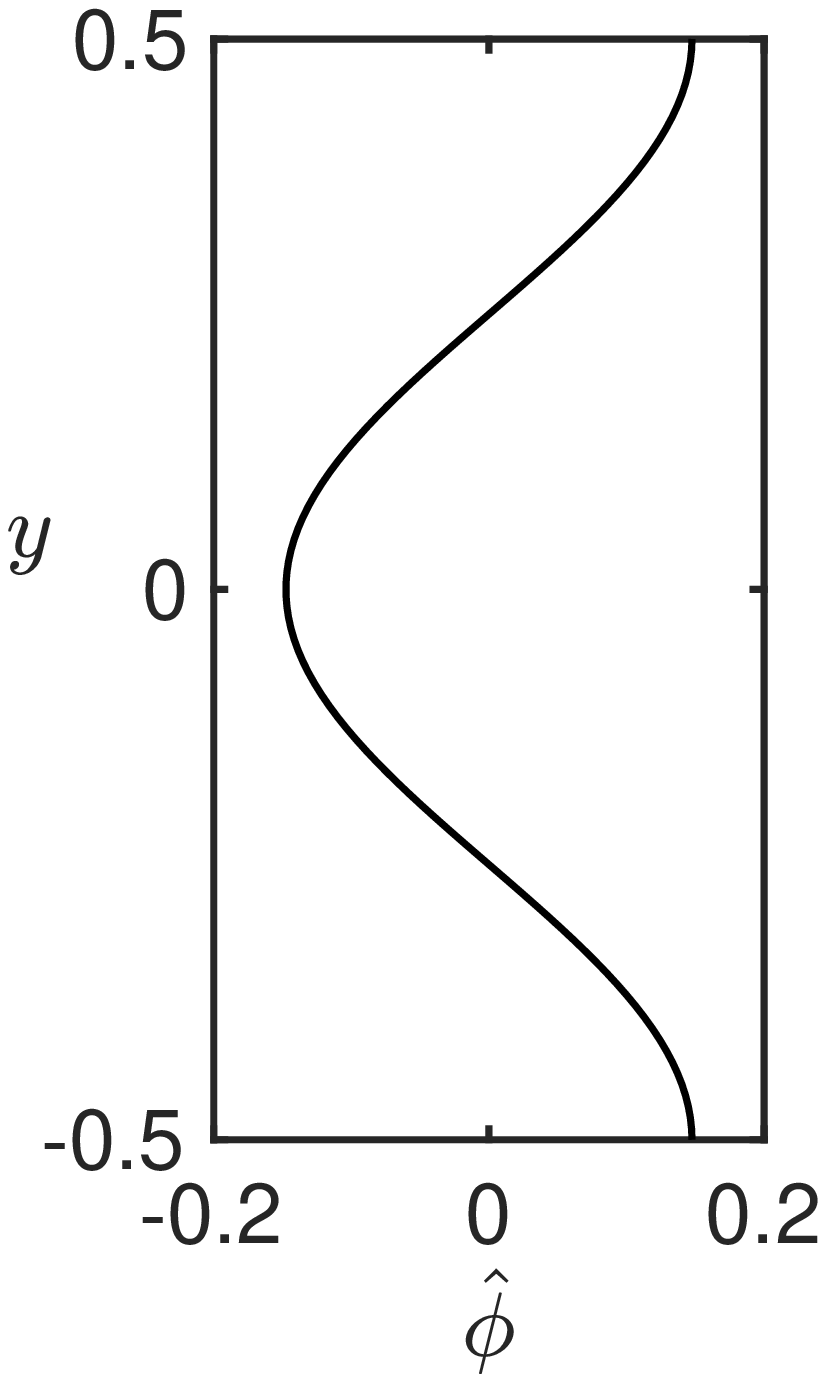}
\end{minipage}
\caption{Qualitative comparison of the linear stability theory with the molecular dynamics simulation for parameter values $e=0.8$, $H=80$, $\phi^0=0.6$: (left) particle position plot of a shear band from~\cite{CG2004} and (right) eigenfunction of density $\hat{\phi}$ for $\beta=2$.   
}
\label{fig_comparision}
\end{figure}

Figure~\ref{fig_comparision} compares \emph{qualitatively} a typical result from molecular dynamics simulation~\cite{CG2004} with that from the present linear stability theory for parameter values $e=0.8$, $H=80$ and $\phi^0=0.6$.  
The left panel in the figure shows two parallel, high-density regions located on both sides of the centerline~\cite{CG2004}. 
The right panel in the figure exhibits the density eigenfunction $\hat{\phi}$ for the same parameter values and for $\beta=2$. 
Similarly to the molecular dynamics simulation result~\cite{CG2004}, the density eigenfunction shows minimum density at the centerline ($y=0$) of the channel and higher densities on both sides of the centerline.   
For these parameter values, the corresponding growth rate is positive showing the instability of the USF. 

\subsection{
Effect of the restitution coefficient and channel width on the shear banding: a global criterion
}
\label{subsec:onset}

\begin{figure}[!ht]
\includegraphics[scale=0.43]
{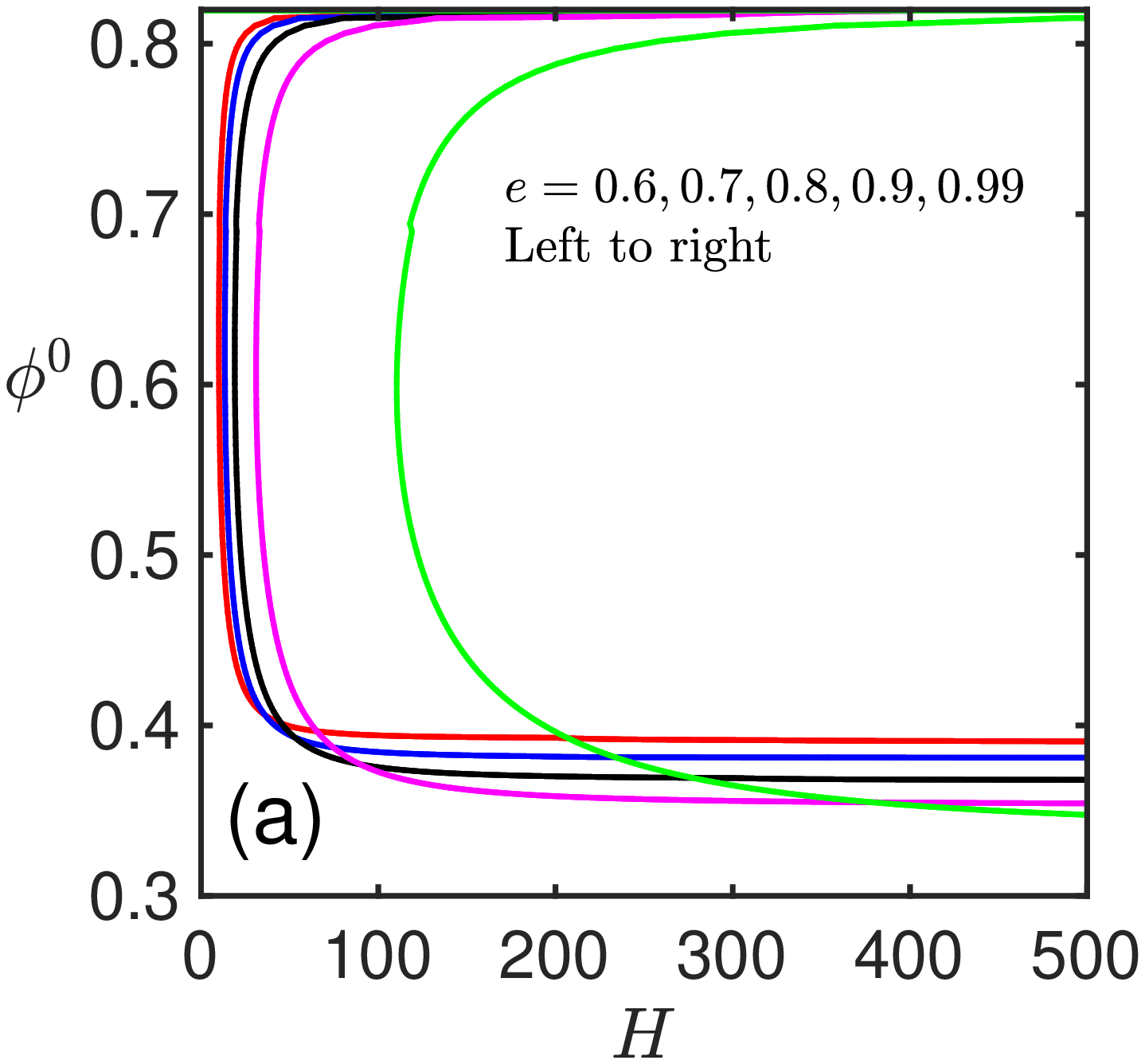}
\qquad\quad
\includegraphics[scale=0.42]
{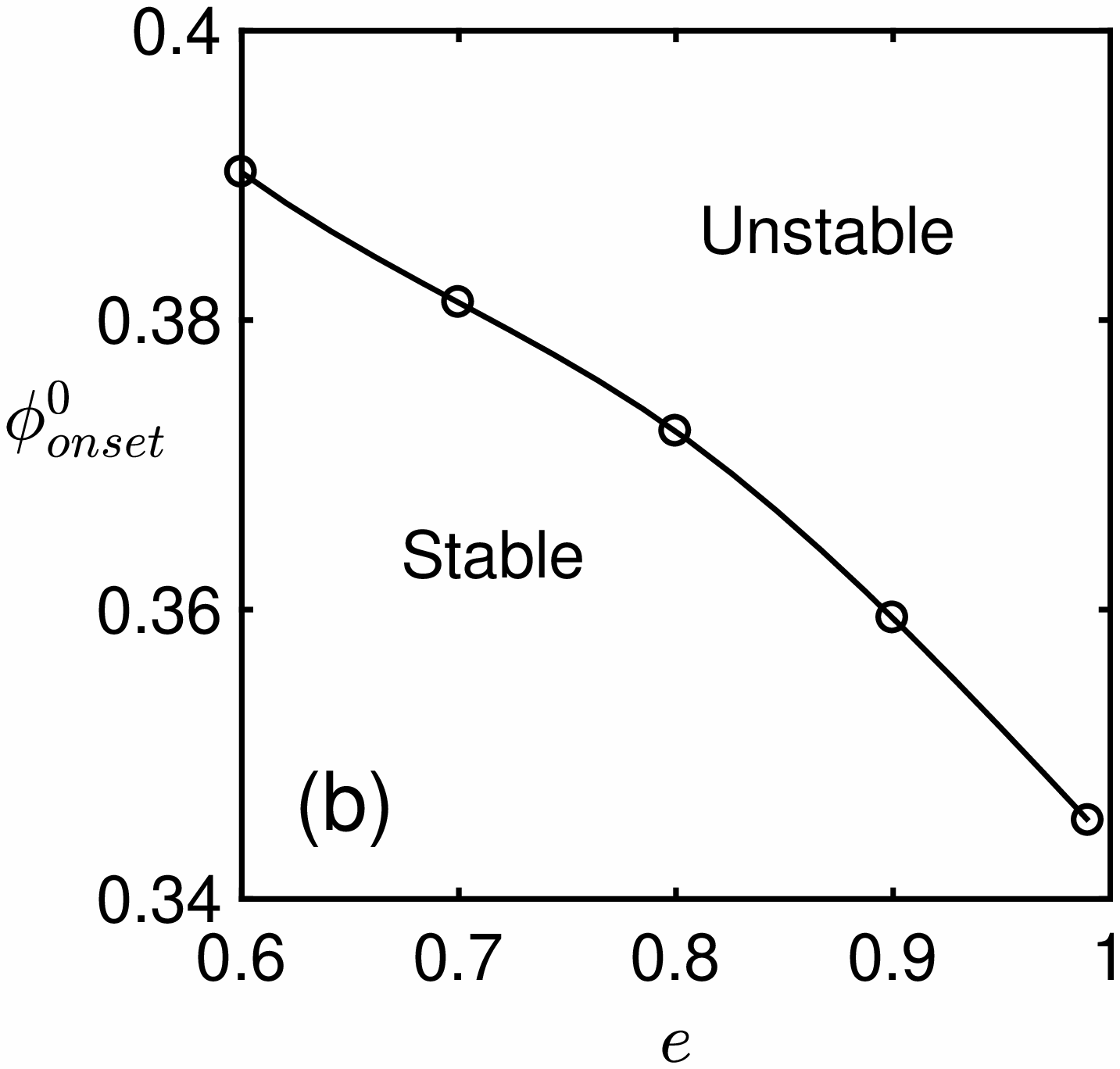}
\caption{(a) Neutral stability curves in the $(H,\phi^0)$-plane for increasing restitution coefficient $e=0.6,  0.7, 0.8, 0.9, 0.99$; the flow is stable outside each neutral stability curve and unstable inside it. 
(b) Variation of $\phi^0_{onset}$ with $e$ where the flow is unstable for $\phi^0>\phi^0_{onset}$ for each $e$.}
\label{fig:effect_restitution}
\end{figure}

%
Figure~\ref{fig:effect_restitution}(a) illustrates the neutral stability curves in the $(H,\phi^0)$-plane for various restitution coefficients ranging from moderately inelastic to quasi-elastic limit. For a fixed restitution coefficient, the USF is unstable inside each contour and stable outside. 
Similarly to figure~\ref{fig1:contours_ep5}(a), figure~\ref{fig:effect_restitution}(a) also shows that the USF is stable in the dilute limit ($\phi^0 \to 0$) for all values of the restitution coefficient, which is in agreement with previous studies, e.g.~\cite{AN1998, Garzo2006, ASL2008}.
We see that the neutral stability curve shifts towards right as the restitution coefficient increases thereby leading to more stable region behind it. In other words, the range of the channel width for which the USF is unstable decreases with increasing the restitution coefficient. Indeed, there exists a set of critical parameters ($\phi^0_{c}, H_{c}$) corresponding to the boundary of a neutral stability curve outside which the USF becomes unstable.

As discussed earlier, the lower branch of each neutral stability curve in figure~\ref{fig:effect_restitution}(a) asymptotically (as $H\rightarrow \infty$) approaches to a minimum mean density
below which the flow is always stable. 
We define this minimum critical density as the onset density, i.e.%
\begin{equation}
\label{eqn:onset_density}
\phi_{onset}^0(e) := \min_{H}\phi_{c}^0(H,e)\quad\mbox{as}\quad H\to \infty,
\end{equation}
where the USF is stable for $\phi^0<\phi^0_{onset}$ and vice versa.
The onset mean density~\eqref{eqn:onset_density} as a function of the restitution coefficient is plotted
in figure~\ref{fig:effect_restitution}(b). 
It is clear that the onset density decreases with increasing the restitution coefficient, which implies that the USF tends to lose  stability at a lower density in the elastic limit than that in the inelastic case. 
Note that the above conclusion holds for very large channel widths. 



\begin{figure}[!b]
\includegraphics[scale=0.46]
{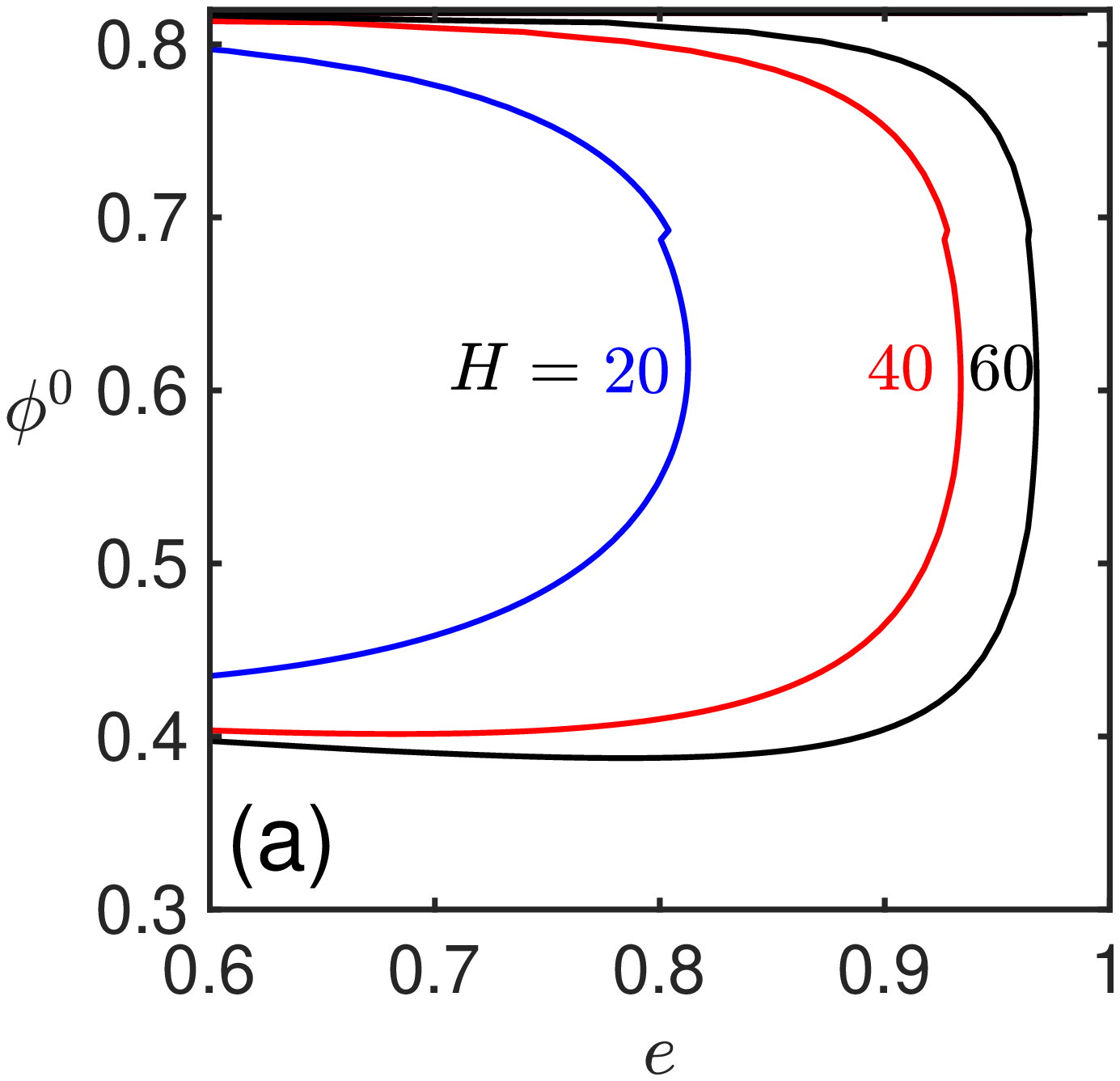}
\quad\qquad
\includegraphics[scale=0.45]{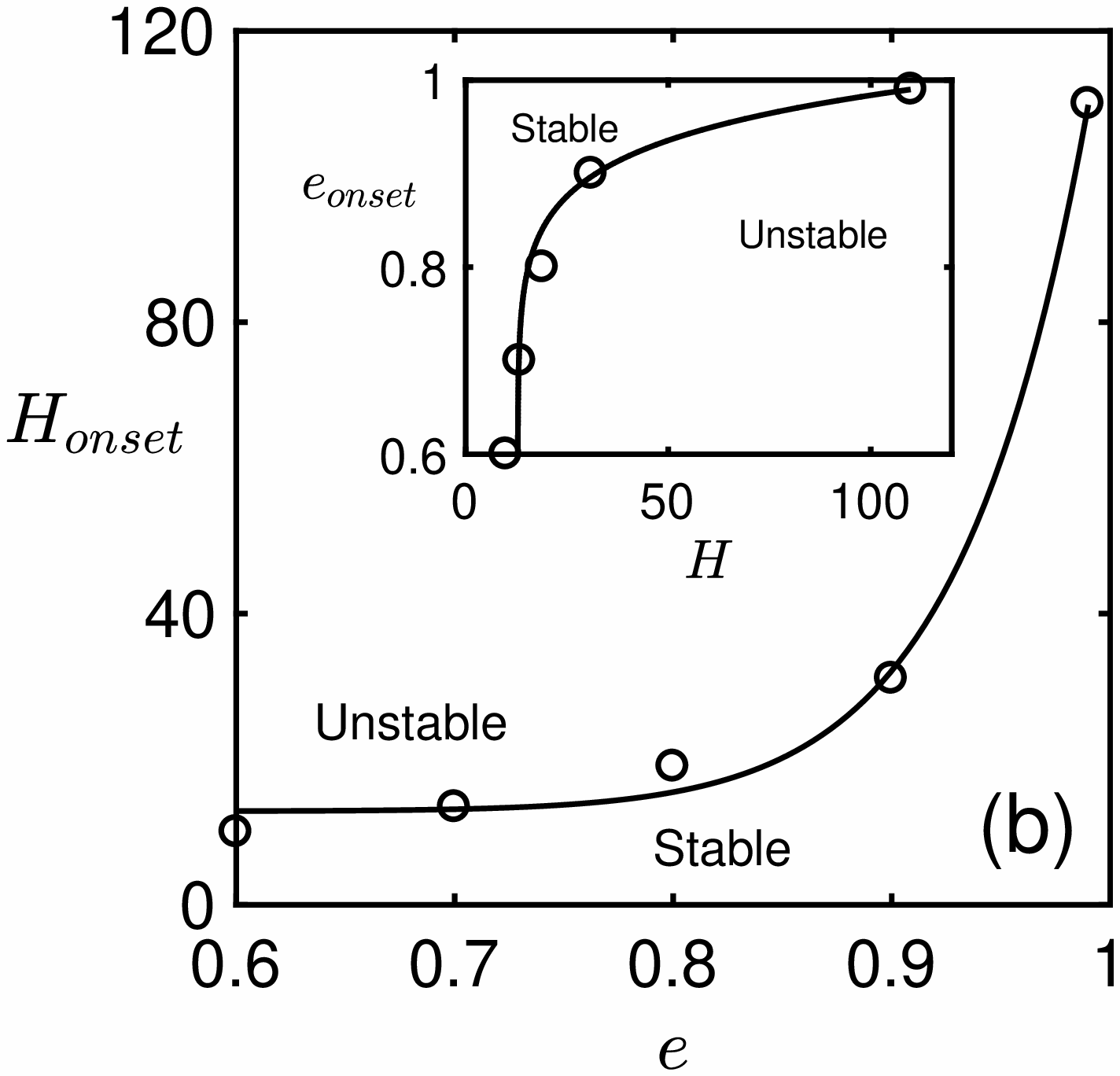}
\caption{(a) Neutral stability curves in the $(e,\phi^0)$-plane for increasing channel width $H=20$, $40$, $80$,
 and
(b) variation of $H_{onset}$ with $e$ (main panel), and $e_{onset}$ with $H$ (inset), where the solid line and circles represent the values from relation~\eqref{eqn:onsetH_e} and those extracted from the neutral stability curves in figure~\ref{fig:effect_restitution}, respectively.}
\label{fig:effect_restitution1}
\end{figure}

It can also be noticed 
from figure~\ref{fig:effect_restitution}(a) that there exists an onset value of the channel width---associated with the nose of each neutral stability curve---
below which the USF is always stable for all mean densities and above which it is unstable.  
Let us define this onset value of the channel width as the minimum of all critical channel widths:
\begin{equation}
\label{eqn:honset_e}
H_{onset}(e) := \min_{\phi^0} H_{c}(\phi^0,e).
\end{equation}
Note that $\phi_{c}^0$ and $H_{c}$ depend on both the mean density and restitution coefficient whereas $\phi_{onset}^0$ and $H_{onset}$ are only functions of the restitution coefficient and hence of the inelasticity.  
It is evident from figure~\ref{fig:effect_restitution}(a) that the value of $H_{onset}$ increases with increasing the restitution coefficient implying that the USF is more stable as $e$ increases which qualitatively matches with the simulation results of~\cite{CG2004,CLG2006}. Owing to the functional dependence (see~\eqref{eqn:honset_e}), the shear banding in granular shear flow starts to appear at small channel widths when particles are more inelastic as compared to the quasi-elastic ones. 
In contrast, the granular shear flow 
remains uniform for large values of the channel widths.


In order to get further insight, the neutral stability curves in the $(e,\phi^0)$-plane are shown for three values of the channel width in figure~\ref{fig:effect_restitution1}(a). 
The flow is stable outside (towards right) of each contour and unstable inside the bounded region of the curve (towards left). 
As the channel width increases, the neutral stability curve shifts towards right such that it covers more unstable region in the $(e, \phi^0)$-plane, i.e.~the flow becomes more unstable with the increasing channel width.  
Therefore for a fixed density, 
the range of the restitution coefficient for which the USF is unstable increases with increasing channel width.  
Figure~\ref{fig:effect_restitution1}(a)
also depicts that there is an onset restitution coefficient below which the shear-banding instability persists and above which the USF remains stable. 
We define this onset restitution coefficient as:
\begin{equation}
e_{onset}(H) := \max_{\phi^0} e_{c}(\phi^0,H).
\end{equation}
It can also be seen from the inset of figure~\ref{fig:effect_restitution1}(b) that $e_{onset}$ increases with increasing the channel width. 

Although the neutral stability curves shown in figures~\ref{fig:effect_restitution} and~\ref{fig:effect_restitution1} depict the overall behavior of the instability, it still remains to understand 
how the onset parameters $H_{onset}(e)$ and $e_{onset}(H)$ vary  
and to discern if there exists any relation relating these parameters. 
%
%
%
We shall now seek the onset of the shear-banding instability in terms of these onset parameters. 
As mentioned earlier, 
the USF is unstable 
for all $H>H_{onset}$ 
and for all $e<e_{onset}$, 
and
%
these onset values correspond to the nose of each neutral stability curve, see figure~\ref{fig:effect_restitution1}(a). 
%
%
The onset parameters
$H_{onset}(e)$ (main panel) and $e_{onset}(H)$ (inset) are shown for some points by circles in figure~\ref{fig:effect_restitution1}(b).  
By curve fitting, 
one can find a functional relationship between the onset parameters $e_{onset}$ and $H_{onset}$, which reads
\begin{equation}
\label{eqn:onsetH_e}
H_{onset}= \alpha e_{onset}^\gamma + \delta,
\end{equation}
where $\alpha=115$, $\gamma=17$ and $\delta=12.86$. 
The solid line in figure~\ref{fig:effect_restitution1}(b) represents the values obtained from relation~\eqref{eqn:onsetH_e}.
It is important to note that~\eqref{eqn:onsetH_e} is a global criterion as this does not depend on the  
spatial positions.





%

\begin{figure}[!ht]
\vspace{1em}
\includegraphics[scale=0.45]
{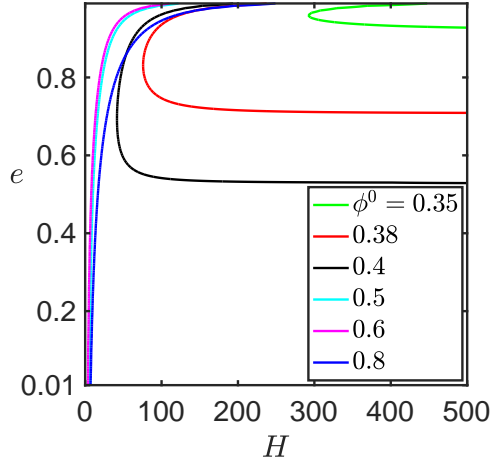}
\caption{Neutral stability curves in the $(H,e)$-plane for different values of $\phi^0$. The flow is stable outside (left) of each contour and unstable inside it (right).}
\label{fig2:effect_restitution}
\end{figure}

Let us now analyze the effect of the mean density on the shear-banding instability. 
Figure~\ref{fig2:effect_restitution} illustrates the neutral stability curves in the $(H,e)$-plane for various mean densities.  
For a fixed mean density, the USF 
is unstable inside (towards right of) each of the neutral stability curves. 
The neutral stability curves for less dense flows look markedly different from those of  moderately-to-highly dense flows.  
%
%
It is also seen that the unstable region increases with increasing the mean density, therefore the shear-banding instabilities are more prone to the dense 
flows, in general. However, for densities $\phi^0 > \phi_f \approx 0.69$, the instability region decreases with increasing densities, see figure~\ref{fig2:effect_restitution}.  

\subsection{
Effect of the restitution coefficient and density on the shear banding: another global criterion}
\label{subsec:onset_phi}

As discussed above, the shear-banding instability is stationary which implies that 
the least stable mode, and hence the dominant mode, is real. 
Therefore the onset of the shear-banding instability, where the growth rate is zero, can be determined analytically from dispersion relation~\eqref{DISPREL} by substituting $\omega=0$ which gives $a_0=0$. 
Using the expression of $a_0$ from~\eqref{ai} (see appendix~A), one obtains  
\begin{eqnarray}
\frac{\pi^2 \beta^2}{H^2} = 
\frac{\xi_1^0}{\xi_2^0},
%
%
\label{eqn:criteria}
\end{eqnarray}
where
\begin{equation}
\xi_1^0 = \left( \frac{f_{5\phi}^0}{f_5^0} + \frac{f_{2\phi}^0}{f_2^0}
\right)\frac{f_1^0}{f_{1\phi}^0}  -2
\quad \mbox{and}\quad
\xi_2^0 =
\frac{1}{f_5^0}  \left( f_4^0 - 
\frac{f_1^0}{f_{1\phi}^0}
f_{4h}^0\right).
\end{equation}
Thus  at the onset of the shear-banding instability, equality~\eqref{eqn:criteria} must be satisfied. 
Note that the left-hand side of~\eqref{eqn:criteria} is always positive and therefore
\begin{align}
\frac{\xi_1^0}{\xi_2^0}>0 
\quad\implies\quad 
\xi_1^0>0,
\end{align} 
because $\xi_2^0>0$.
 Following~\cite{Alam2006,ASL2008}, the condition $\xi_1^0>0$ is equivalent to
\begin{align}
\frac{\partial }{\partial \phi^0} \left[ \frac{\sqrt{f_2^0 f_5^0}}{f_1^0}\right] >0 \quad \textrm{provided} \quad 
\frac{\partial f_1^0}{\partial \phi^0}>0,  
\label{eqn:cond1}
\end{align}
which must be satisfied 
at the onset of instability. 
Thus condition~\eqref{eqn:cond1} leads to a necessary criterion for the shear-banding instability. 
It is worth noticing that the term $\sqrt{f_2^0 f_5^0}/f_1^0$ is none other than the ratio of the shear stress to the pressure of the USF. 
This further allows us
to express the dynamic friction coefficient, the ratio of 
sliding force $F_s$ to normal force $F_n$, as
\begin{align}
C_f  \equiv C_f(\phi^0,e) = \frac{F_s}{F_n} = \frac{\eta^0  u^0_y  }{p^0} = 
\frac{\sqrt{f_2^0 f_5^0}}{f_1^0}.
\label{eqn:cond2}
\end{align}
Here we have used the definition of the base state granular temperature $T^0=f_2^0/f_5^0$ and of the base state shear rate $u^0_y=1$, see Sec.~\ref{subsec:basestate}.
From~\eqref{eqn:cond1} and~\eqref{eqn:cond2}, it can be concluded that 
the existence of the shear-banding instability
requires the dynamic friction coefficient $C_f$ to be an increasing function of mean density $\phi^0$, i.e.~$\partial C_f/\partial \phi^0>0$.  
In other words, the condition for the onset of the shear-banding instability is given by
\begin{equation}
\lim_{\phi^0\to \phi^{0+}_{onset}} \frac{\partial C_f}{\partial \phi^0}=0.
\label{cond}
\end{equation} 
Note that condition~\eqref{cond} 
is also a global criterion for the onset of the shear-banding instability as it also does not depend on the spatial positions.    

\begin{figure}[!htbp]
\includegraphics[scale=0.45]
{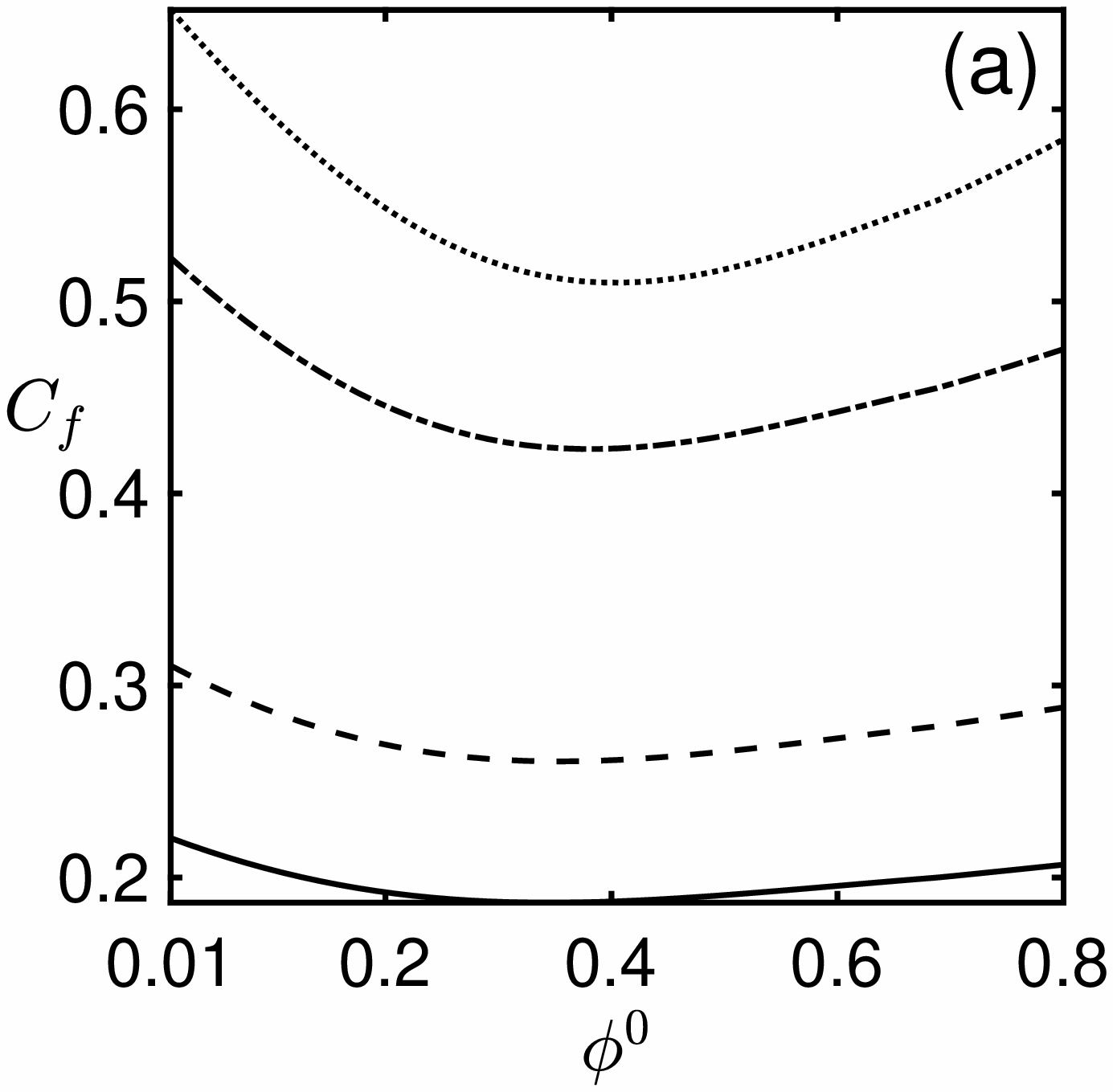}
\qquad\qquad
\includegraphics[scale=0.45]
{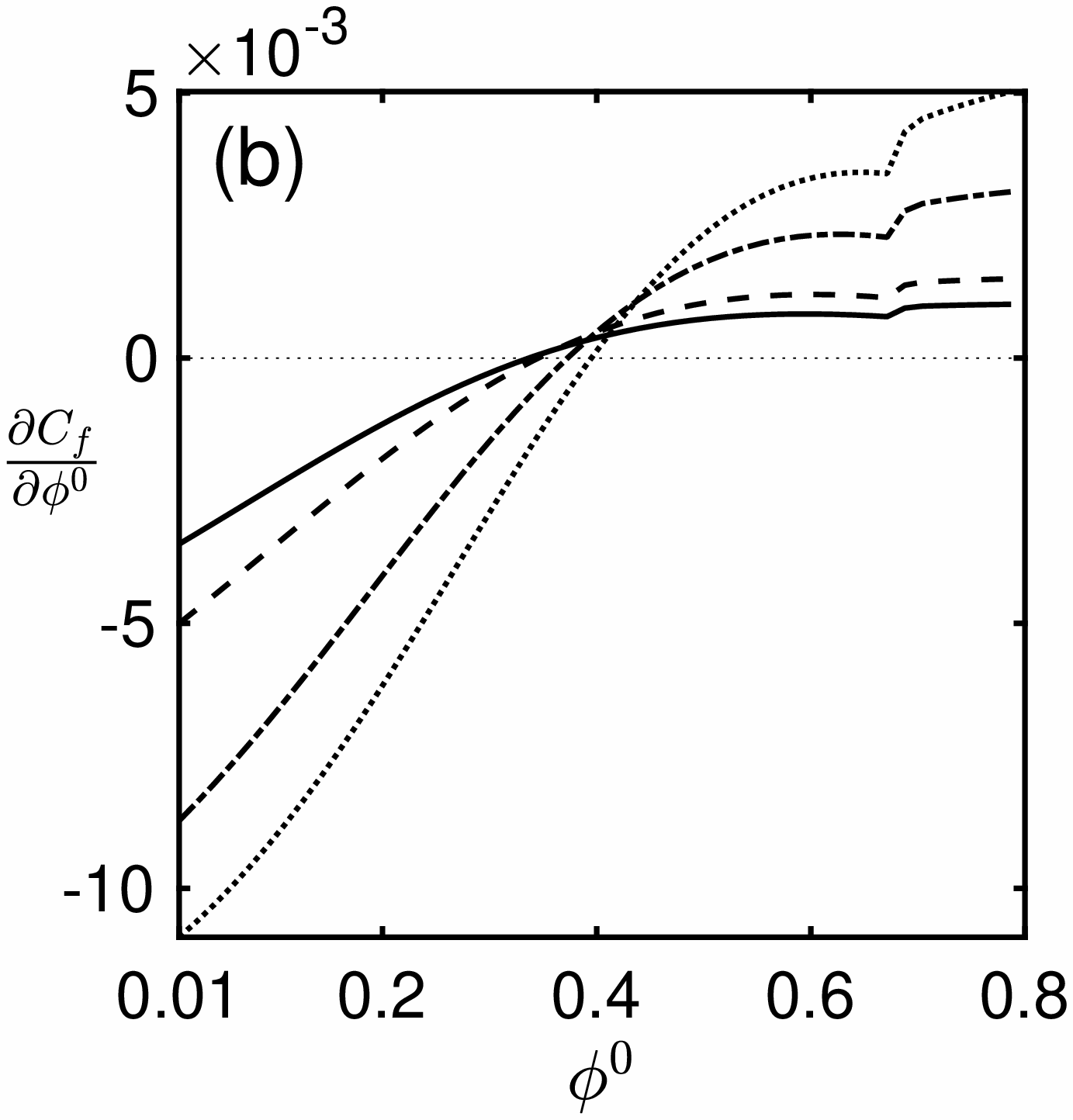}
\caption{Variation of (a) the dynamic friction coefficient $C_f$ and (b) $\partial C_f/ \partial \phi^0$ with
the mean density for various restitution coefficient $e$: 0.5 (dotted line), 0.7 (dash-dotted line), 0.9 (dashed line) and 0.95 (solid line). 
}
\label{fig:cf}
\end{figure}

Figure~\ref{fig:cf} shows the variation of the dynamic friction coefficient $C_f$ and its gradient with respect to the mean density $\phi^0$ for various restitution coefficients. 
It is seen from figure~\ref{fig:cf}(a) that the dynamic friction coefficient $C_f$ varies non-monotonically---it first decreases, attains a minimum, and increases thereafter with increasing density. 
A value of the mean density $\phi^0$ where $C_f$ attains its minimum is the same as the onset density $\phi_{onset}^0$ 
in figure~\ref{fig:effect_restitution1}(a) (for corresponding $e$),
which was extracted from the neutral stability curve. 
Clearly, at $\phi^0=\phi_{onset}^0$, 
the slope $\partial C_f/ \partial \phi^0$ is zero, see figure~\ref{fig:cf}(b). 
 
\begin{figure}[!bt]
\includegraphics[scale=0.43]
{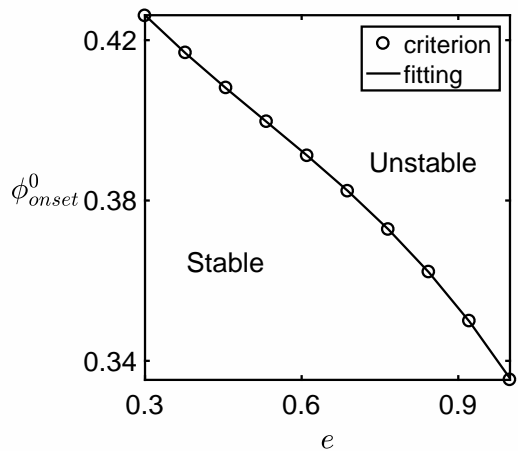}
\caption{Variation of $\phi^0_{onset}$ with $e$. 
The solid lines are the values computed from~\eqref{eqn:rel_onsetphi_e} and circles represent the values obtained from criterion~\eqref{cond}.}
\label{fig:onset_phi_global}
\end{figure}

Similarly to figure~\ref{fig:effect_restitution}(b),
figure~\ref{fig:onset_phi_global} also depicts the variation of the onset mean density $\phi_{onset}^0$ with the restitution coefficient calculated using criterion~\eqref{cond} (solid line). 
Clearly, the onset mean density decreases monotonically with increasing $e$, which implies that the onset of the shear-banding instability in the elastic limit ($e \to 1$) occurs at slightly lower mean density than in the case of restitution coefficient $e<1$.
%
%
By curve fitting, 
one can find a functional relationship between
$\phi_{onset}^0$ and $e$, which reads 
\begin{align}
\phi_{onset}^0 = \alpha_1 e^3 + \alpha_2 e^2 +\alpha_3 e + \alpha_4, 
\label{eqn:rel_onsetphi_e}
\end{align}
where $\alpha_1=-0.1350$, $\alpha_2=0.2144$, $\alpha_3=-0.2208$ and $\alpha_4= 0.4768$. 
This implies that criterion~\eqref{cond} for the onset of the shear-banding instability is equivalent to relation~\eqref{eqn:rel_onsetphi_e}.
 

\section{Conclusions and outlook}
\label{sec:Conclusion}
In this paper, we have extended 
a previous study~\cite{ASL2008} and
investigated the shear-banding instability in an arbitrary inelastic granular shear flow. In particular, we have analyzed the stability of the USF using granular hydrodynamic equations closed with the Navier--Stokes-level constitutive relations proposed by~\citet{GD1999} and~\citet{Lutsko2005}. 
Two limiting cases of the full GDL model, namely the kinetic limit, which is valid for dilute flows, and the collisional limit have also been discussed. 
It has been shown that the 
kinetic limit of the GDL model is able to give correct prediction about the stability of dilute granular shear flows. 
Surprisingly, the collisional limit alone fails to give meaningful prediction on the instability of dense granular shear flows. 
Thus we have shown that both the kinetic and collisional mechanisms are important to capture
the shear-banding instability in dense granular shear flows correctly.  
Furthermore, by analyzing the neutral stability curves for various parameters in different planes, we have found that the USF is always stable in the dilute limit and therefore non-uniformity in terms of shear band does not appear in dilute granular flows. 
In contrast to dilute granular shear flows, moderately-to-highly dense granular shear flows become unstable once the control parameters exceed their critical values.   

The influence of the restitution coefficient, channel width and mean density on the shear banding has also been explored. 
It has been found that the USF is more unstable with: (i) decreasing the restitution coefficient (or increasing the inelasticity) for any fixed mean density and fixed channel width, (ii) increasing the channel width for any fixed mean density and fixed restitution coefficient, and (iii) increasing the mean density (in general) for any fixed channel width and fixed restitution coefficient. 
These findings all together lead to the fact that the shear banding is more prone to dense flows of highly inelastic particles confined in channels having large channel widths. 
This fact agrees {\it qualitatively} with the simulation results of~\citet{CG2004,CLG2006},  
who showed that a pseudo one-dimensional cluster appears with increasing the restitution coefficient or with increasing the channel width 
while fixing the other parameters,
and therefore the intensity of the clustering instability can be 
controlled 
with these parameters.

In the present study, the onset values of the density and channel width 
as a function of the restitution coefficient have been assessed. For this, we have  defined three parameters:  
(i) $\phi^0_{onset}(e)$, the smallest mean density below which the USF is always stable for all channel widths and above which it is unstable, (ii) $H_{onset}(e)$, the critical channel width below which the flow is stable for all mean densities and above which it becomes unstable and (iii) $e_{onset}(H)$, the critical restitution coefficient above which the flow is always stable for all mean densities and below which it is unstable. 
Furthermore, a relation between $H_{onset}$ and $e_{onset}$ has been obtained in a power-law form. From this relation, one can easily find
the onset channel width for the shear-banding instability at a fixed restitution coefficient and vice versa.  
This is a global criterion (as it does not depend on spatial locations) for the
onset of the shear banding in terms of $e$ and $H$---the shear banding appears for $H>H_{onset}(e)$ or for $e<e_{onset}(H)$. 

It has been found that
the onset of the shear-banding instability in 
granular USF
is tied to the increasing dynamic friction coefficient, $C_f(\phi^0,e) = \eta^0 u_y^0/p^0$ with $\eta^0$ and $p^0$ being the shear viscosity and pressure, respectively, in the USF. 
In particular, the USF breaks into the dense and dilute regions of low and high shear (shear stress or shear rate) along the gradient direction 
when 
the dynamic friction coefficient 
increases with the mean density.  
In other words, the USF cannot sustain higher friction with increasing the mean density,  
and therefore rearranges
to 
a non-uniform shear-banded state of lower dynamic friction.    
For a fixed restitution coefficient, we have found that the gradient of the dynamic friction coefficient with respect to the mean density takes values from negative to positive and crosses zero at the onset mean density   
$\phi_{onset}^0$. 
%
The onset mean density has been found to be a monotonically decreasing function of the restitution coefficient satisfying the
cubic polynomial relation~\eqref{eqn:rel_onsetphi_e}. 
Consequently, the USF of nearly elastic particles reaches to the onset of the shear-banding instability at a lower mean density as compared to that of relatively more inelastic particles, which 
is in contrast to~\cite{ASL2008} as the constitutive models employed in~\cite{ASL2008} are valid only for nearly elastic particles. 
Furthermore, 
the onset of the shear banding 
has been found to follow 
two global criteria relating  
(i) the channel width and restitution coefficient and
(ii) the mean density and restitution coefficient (or the shear viscosity and pressure).

Within the framework of dense   granular flows with arbitrary inelasticity, the present 
work provides 
the control parameters for the
onset of the shear-banding instability.
Functional relationships relating the onset parameters 
have been obtained that enable us to predict the existence of shear bands in a granular shear flow by merely knowing the control parameters. 
This could also be of much interest from an experimental point of view as the onset of the shear-banding instability becomes completely known from the present work
in terms of control parameters. 
It is important, however, to note that the findings of the present paper are based on the NSF equations closed with the GDL model
that neglects the anisotropy of the USF. 
To incorporate the effect of anisotropy of the USF, the generalized transport coefficients for dense granular flows are required. 
The linear stability analysis of the USF with the generalized transport coefficients will be a topic of  future research. 
Notwithstanding, the present work paves the way for simulations and experiments on granular shear flows
of arbitrarily inelastic particles. %


\section*{Acknowledgment}
The authors acknowledge the anonymous reviewers for their valuable suggestions, which significantly improved the paper. PS acknowledges financial support from IIT Madras through the grant~MAT/16-17/671/NFSC/PRIY. 
VKG gratefully acknowledges the financial supports through the “MATRICS” project MTR/2017/000693 funded by the SERB, India and that through the Commonwealth Rutherford Fellowship.

\appendix

\appendix
\section{Coefficients in dispersion relation \texorpdfstring{\eqref{DISPREL}}{}}
\label{dis_coeff}
The coefficients $a_{ij}$ in \eqref{ai} that enters dispersion relation \eqref{DISPREL} are as follows.
\begingroup
\allowdisplaybreaks
\begin{align*}
a_{30} &= \frac{1}{\phi_0} \left(\mathcal{D}_{0,T}^0 - \eta_T^0\right),
\\
a_{32} &= \frac{1}{\phi_0} \left(3 \eta_0+\kappa_0+\lambda_0\right) \pi^2 \beta^2,
\\
a_{22} &= \frac{1}{\phi_0^2} \left[2 \eta_0 \eta_T^0 +p_0 p_T^0 +\phi_0^2 \, p_\phi^0 + p_T^0 \,\mathcal{D}_1^0  + (3 \eta_0+\lambda_0) \left(\mathcal{D}_{0,T}^0 - \eta_T^0\right) \right] \pi^2 \beta^2,
\\
a_{24} &= \frac{1}{\phi_0^2} \left[ (3 \eta_0+\lambda_0) \kappa_0 + (2 \eta_0+\lambda_0) \eta_0\right] \pi^4 \beta^4,
\\
a_{12} &= \frac{2 p_T^0 \eta_0}{\phi_0^2} \pi^2 \beta^2 
+ \frac{1}{\phi_0} \left[p_\phi^0 \left(\mathcal{D}_{0,T}^0 - \eta_T^0\right) - p_T^0 \left(\mathcal{D}_{0,\phi}^0 - \eta_\phi^0\right) \right] \pi^2 \beta^2,
\\
a_{14} &= \frac{\eta_0}{\phi_0^3} \left[(2\eta_0 + \lambda_0)\left(\mathcal{D}_{0,T}^0 + \eta_T^0\right) + p_0 p_T^0 + p_T^0  \mathcal{D}_1^0\right] \pi^4 \beta^4
+\frac{1}{\phi_0} \left(p_\phi^0 \kappa_0 + p_\phi^0 \eta_0 - p_T^0 \mu_0 \right)  \pi^4 \beta^4,
\\
a_{16} &= \frac{\eta_0}{\phi_0^3} (2 \eta_0+\lambda_0) \kappa_0 \pi^6 \beta^6,
\\
a_{04} &=  \frac{\eta_0}{\phi_0^2} \left[p_\phi^0 \left(\mathcal{D}_{0,T}^0 + \eta_T^0\right) - p_T^0 \left(\mathcal{D}_{0,\phi}^0 + \eta_\phi^0\right) \right] \pi^4 \beta^4,
\\
a_{06} &=  \frac{\eta_0}{\phi_0^2} \left(p_\phi^0 \kappa_0 - p_T^0 \mu_0\right)  \pi^6 \beta^6.
\end{align*}
\endgroup
\bibliography{references}

\end{document}